\documentclass[aps,nofootinbib]{revtex4}

\usepackage{tikz}

\usepackage{amsmath,bbm}
\usepackage{amssymb}
\usepackage{mathrsfs}
\usepackage{mathtools}
\usepackage{amsthm}
\usepackage{xcolor}
\usepackage{verbatim}
\usepackage{bm}
\usepackage{subfigure}
\usepackage{hyperref}
\hypersetup{
	colorlinks=true,
	linkcolor=blue,
	filecolor=magenta,      
	urlcolor=cyan,
}
\newcommand{\R}{\mathbb{R}}

\newcommand{\be}{\nopagebreak[3]\begin{equation}}
\newcommand{\ee}{\end{equation}}
\newcommand{\ba}{\nopagebreak[3]\begin{eqnarray}}
\newcommand{\ea}{\end{eqnarray}}
\DeclareFontFamily{U}{rsfs}{}         
\DeclareFontShape{U}{rsfs}{m}{n}{<5> rsfs5 <6><7> rsfs7          %
  <8><9><10><10.95><12><14.4><17.28><20.74><24.88> rsfs10}{}     %
\DeclareMathAlphabet{\mathfs}{U}{rsfs}{m}{n}                     %
\newcommand{\mfs}[1]{\mathfs {#1}}                               %
\newcommand{\n}{{\nonumber}}

\newcommand{\sD}{{\mfs D}}

\newcommand{\sE}{{\mfs E}}
\newcommand{\sH}{{\mfs H}}
\newcommand{\sL}{{\mfs L}}

\newcommand{\sM}{{\mfs M}}

\newcommand{\sO}{{\mfs O}}

 \usepackage{braket}

\newcommand{\va}{\scriptscriptstyle}

\usepackage{braket}

\begin{document}

\title{The problem of time: a path integral view}

\author{Juan Manuel Diaz}
\email{diaz@cpt.univ-mrs.fr}
\affiliation{Aix Marseille Universit\'e, Universit\'e de Toulon, CNRS, Centre de Physique Th\'eorique, 13000 Marseille, France.}

\author{Alejandro Perez}
\email{perez@cpt.univ-mrs.fr}
\affiliation{Aix Marseille Universit\'e, Universit\'e de Toulon, CNRS, Centre de Physique Th\'eorique, 13000 Marseille, France.}

\begin{abstract}
We show that the problem of the emergence of time evolution in an otherwise timeless non relativistic closed quantum system---considered as a poor's man model of generally covariant quantum theory---can be approached from the perspective of the path integral representation. As it is often the case with the functional integral approach such a perspective offers a more intuitive account of aspects which can become cumbersome in the operator/state Hilbert space formulation of the problem. 
We show how Schrödinger time evolution emerges from the timeless closed quantum system when a clock degree of freedom is identified and put in a suitable semiclassical {\em good-clock state}. 

Our analysis has a simple consequence that can be extrapolated to the formal path integral framework applied to generally covariant systems with action $S$ (including gravity). In those theories certain transition amplitudes are given by the combination $\exp(i S/\hbar)+\exp(-i S/\hbar)$ rather than expected  `forward propagating' $\exp(i S/\hbar)$. This feature,  known as the {\em cosine problem}, is realized in concrete regularizations of the path integral: as in the spin foam representation defining physical inner product between quantum geometry (spin network) states in loop quantum gravity. Both at the general formal level as well as in concrete regularizations, this apparent difficulty has led some authors to consider it as a problem requiring modification of the basic amplitudes in order to avoid backward propagation.

Our model shows how the {\em cosine problem} is an expected consequence of time reversal invariance of the fundamental dynamics together with the time neutral nature of the boundary states customarily used in computations of transition amplitudes.  When a suitable clock system is identified---and put in a suitable (semiclassical) boundary state, a {\em good-clock state}---it  introduces a time arrow that selects the  `forward propagating' $\exp(i S/\hbar)$, without the need of any modification of the fundamental dynamics. 

The analysis provides a natural handle to understanding the emergence of time under suitable conditions and stresses the fact that, in the canonical formulation, quantum gravity is fundamentally timeless.
\end{abstract}

\maketitle


\section{Introduction}

In this paper we review the problem of time in generally covariant systems and propose an amplitude-based path integral perspective on the issue, which we hope is helpful in discussions about the way spacetime physics might emerge from background-independent quantum gravity approaches.
  
\begin{figure}[h] \centerline{\hspace{0.5cm} \(
\begin{array}{c}
\includegraphics[width=15cm]{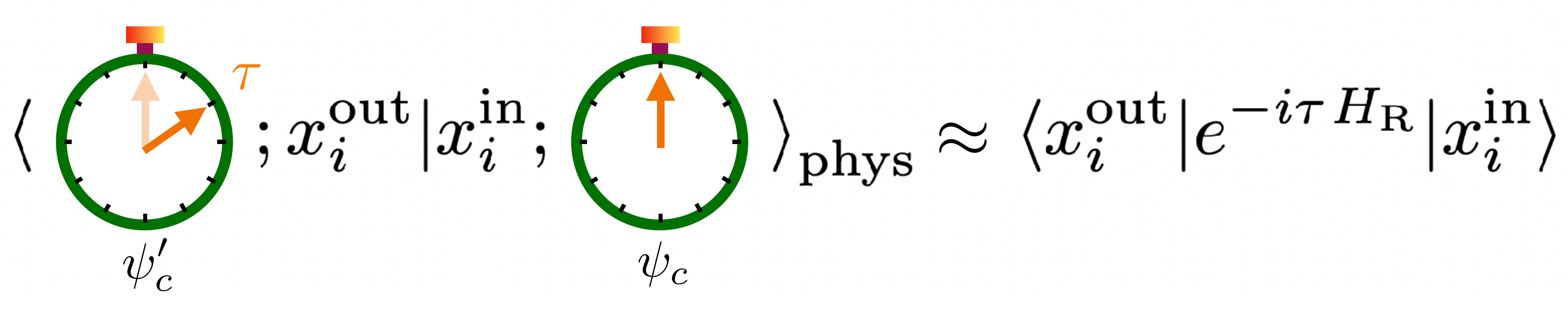} \end{array}
\)}
\caption{{\bf The results in a picture: } In a timeless closed system  where all the allowed quantum states have the same total energy $E$ there is no external time evolution. Assuming that the Hamiltonian is $H=H_c+H_{\rm R}$ where $H_c$ is the Hamiltonian of an internal (clock) degree of freedom, one can write the (boundary) states, for which the transition amplitudes will be computed, as $\bra{\psi_c';x_i^{\rm out}}$ (with $\psi'_c$ and $\psi_c$ denoting the clock states and $x_i$ the rest of the universe  degrees of freedom).
When the boundary clock states are in suitable semiclassical states, their relative features encode internal time $\tau$. In the picture, these states are represented by the clock cartoon inside the Dirac ket and bra to make explicit the fact that they are associated with an internal physical degree of freedom accessible from within the closed system. We will show that the physical transition amplitudes between physical states---computed using the path integral representation---encode the 
`evolution'  of the rest of the universe according to the Schrödinger equation in the emergent time $\tau$.
All the quantities labelling the physical states can be coded in Dirac observables (not involving energy exchanges with anything outside the closed system). As a corollary of the analysis we find that the clock boundary states select the forward propagating amplitude from the 
time neutral path integral amplitude resolving the so-called {\em cosine problem} in the path integral representation of dynamics of generally covariant systems. 
 }
\label{clocky}
\end{figure}

Concrete calculations focus on the emergence of time in the quantum mechanics of a (non relativistic) closed system as a concept defined in relation to a genuine degree of freedom that plays the role of a clock. The system is closed in the strong sense that all allowed quantum states of the system are eigenstates of the total Hamiltonian $H$ with a fixed energy eigenvalue $E$ (a poor man’s model of quantum gravity). When some degree of freedom is associated with a clock and put in a suitable (good-clock) semiclassical state, transition amplitudes in the physical Hilbert space reproduce the unitary Schrödinger evolution in internal clock time (Figure \ref{clocky} is a schematic representation of the main result). Moreover, clock time is encoded in Dirac observables (quantities commuting with $H$) characterizing the clock states.

The subject of time in closed quantum systems has a long history \cite{Kuchar:1991qf}, and the present approach is conceptually tied to the same basic ideas. However, most of the literature studies the issue from the state/operator perspective (see \cite{Hoehn:2019fsy} and references therein). In this paper we take the path integral view and investigate the conditions under which unitary evolution in physical internal time is recovered directly from the transition amplitudes. This makes it easier to treat the problem by identifying realistic clock degrees of freedom (like a free particle, a harmonic oscillator, or otherwise) with bounded-below Hamiltonians $H_c$.

There are important issues concerning the interpretation of quantum theory in a closed system that are intertwined with the problem at hand. We will show that the path integral representation offers an insightful perspective on the problem of time that is complementary to existing analyses, and has the advantage of circumventing, to a certain extent, some of these conceptual difficulties.

Now, such circumventing can be considered a blessing or a curse: it allows for certain potentially useful insights while leaving aside issues that no one knows---at least in an uncontroversial manner---how to deal with.

To give a hint of what we mean, recall that in the standard path integral representation of quantum mechanics we can loosely say that all possible histories of a dynamical system contribute to the dynamics with weights that are equal in magnitude and differ only by a phase controlled by the action of the theory $S$. In this framework, classical trajectories are singled out as those for which the amplitudes contribute coherently and satisfy the extremal condition $\delta S=0$ (linking stationary phase coherence with the classical equations of motion). From this, one may argue that the path integral offers a more intuitive picture of the emergence of classical dynamics—with its clear-cut notion of reality—from the underlying quantum theory. While this is true to a certain extent, the ontological status of quantum theory and the emergence of classical reality still involve important open questions, despite the intuitive insights that the path integral perspective may provide.

In the treatment of this paper, time dynamics in an otherwise timeless closed system emerges in a similar fashion. This does not mean that the conceptual questions previously mentioned are resolved, but some aspects of the ones that concern the emergence of classical dynamics appear to us from what seems a clearer angle.

The previous discussion sets the level at which the problem of time will be approached in this paper. The path integral perspective provides an intuitive, quantum-amplitude-based account of the emergence of time in a closed system, while leaving aside the more difficult interpretational questions of quantum theory that lie beyond the scope of this modest analysis. The issues that the nature of our analysis leaves out (such as aspects of the measurement problem  and the fate---or meaning---of the probabilistic interpretation of quantum theory in a closed system \cite{sep-qt-issues}) represent a significant gap that should not be dismissed.

As mentioned, most concrete calculations are done in the context of a non-relativistic generic quantum system. Nevertheless, the analysis produces nontrivial insights concerning important issues that are currently discussed in the context of the construction of the path integral representation of quantum gravity. Besides the question of the emergence of spacetime from a fundamental background-independent quantum theory, there has been some controversy around what experts in the area of spin foams call the {\em cosine problem}. We believe that, despite its simplicity, our analysis contributes to clarifying this important issue.

The paper is organized as follows. In the next section we review features that make gravity (general relativity or any diffeomorphism-invariant background-independent field theory) a fundamentally closed system for which there is no external notion of time evolution. We recall how and why this is not a serious issue at the classical level (where a spacetime perspective is possible and useful), while it makes quantum physics fundamentally spacetime-less. Such a general discussion should clarify why the key conceptual difficulty associated with the emergence of time can be captured and discussed in a simpler non-relativistic model of a closed system with fixed energy eigenvalue. In Section \ref{333} we briefly review the usual approaches to the problem of time and discuss the case of the functional integral formulation for a non-relativistic toy example in Subsection \ref{pipi}. In Section \ref{coseno} we define the cosine problem in the context of quantum gravity and translate it to the non-relativistic toy model situation. In Section \ref{calculos} we explicitly show the main results of the paper, leading to the cartoon represented in Figure \ref{clocky} and the resolution of the cosine problem. We conclude with a discussion of the results in Section \ref{discu}.

\section{Gravity and general covariance} \label{4d}

The key revolution brought about by general relativity lies less in the specific form of Einstein’s equations and their many physical implications, and more in the profound idea of general covariance. This concept asserts that there are no fixed background fields or structures with respect to which non-dynamical reference systems can be defined. Even though one may rightly argue that general relativity is merely a large-scale, classical approximation to a deeper and still unknown more fundamental theory, general covariance stands out as an inescapable feature of the physical world that will have to persist in the more fundamental formulation that one seeks.

In classical gravity, general covariance implies the absence of any predefined notion of a time variable, which prevents a straightforward interpretation of dynamical evolution---a difficulty commonly referred to as the problem of time. Although this issue is often discussed in the context of quantum gravity, it is important to emphasize that its latent origins lie within the classical theory, general relativity, itself. Considering the tension as already present in the classical framework sheds light on a constructive viewpoint from which the problem of time reveals its true character: not as a puzzle to be solved, but as an intrinsic and unavoidable feature of generally covariant physics. This analysis also provides the conceptual tools for understanding how time can emerge under suitable conditions.  We will also see that the discussion is intimately related to what seems to be more technical issue arising in the path-integral formulation of quantum gravity; namely, the fact that timeless transition amplitudes are controlled by the (time-arrow-neutral) cosine of the action instead of the usual  forward exponential of it (he cosine problem). We will discuss these aspects in more detail below but first 
let us review the conceptual aspects of general covariance in terms of the classical description of gravity.
 
For simplicity, in this introduction we will describe the problem in the context of vacuum general relativity where dynamics is encoded in the 
Einstein-Hilbert action, namely
\ba\label{action}
S[g_{ab}]&=&\int\limits_{\sM} dx^4 \sqrt{|g|} \, R[g_{ab}] \n \\
&=& \int\limits_{\sM=\Sigma\times \R} dx^4 N \sqrt{h} \, \left(R^{\va (3)}[h_{ab}]+K_{ab} K^{ab}-K^2\right).
\ea  
In the second line, the Gauss–Codacci relations have been used, assuming that $\sM$ can be foliated in terms of the coordinate $x_0=$constant hypersurfaces with unit normal $n_a$, so that $h_{ab}=g_{ab}+n_a n_b$ and the extrinsic curvature is given by $K_{ab}=\sL_n(h_{ab})$. The vector field $\partial^a_0$ is decomposed as $\partial^a_0=N n^a+N^a$, where $N$ and $N^a$ are the lapse and shift, respectively. Finally, $R^{\va (3)}[h_{ab}]$ denotes the scalar curvature of the spatial slices \cite{Wald:1984rg}. Denoting derivatives with respect to the coordinate $x^0$ with a dot, one can complete the Legendre transformation and write the previous action in Hamiltonian form as follows
\ba\label{caction}
S[N, N^a,h_{ab}, \pi^{ab}]&=&\int dx^0\int\limits_{\Sigma} dx^3\dot h_{ab} \pi^{ab}-\underbrace{N\sqrt{h}\left(-R^{\va (3)}+\frac{\pi^{ab}\pi_{ab}-\frac12 \pi^2}{h} \right)+2 N^a \nabla^{b}_{\va (3)}\left(\frac{\pi_{ab}}{\sqrt{h}}\right)}_{\equiv H(N, N^a, h_{ab}, \pi^{ab})},
\ea
with the induced metric $h_{ab}$  and conjugate momentum density $\pi^{ab}$ phase space variables with Poisson brackets
\be
\left\{h_{ab}(x^0,x^i)^{},\pi^{cd}(x^0,x^i)^{}\right\}=\delta_{(a}^{c}\delta_{b)}^{d} \delta^{(3)}( x^i, y^i).
\ee  
The variables  $N$, and $N^a$ appear as Lagrange multipliers and the Hamiltonian density of gravity turns out to be
a linear combination of constraints. More precisely, stationarity of the action with respect to $N,N^a, h_{ab}$ and $\pi_{ab}$ yields 
Einstein's equations in Hamiltonian form, which are given by the Hamiltonian constraint defined in \eqref{caction}
\be\label{hamihami}
H[N,N^a,\pi^{ab},h_{ab}]=0,
\ee
for all possible choices of fields $N(x^0,\vec x)$ and $N^a(x^0, \vec x)$,
together with the `evolution' equations
\be\label{evoevo}
\frac{{\partial h}_{ij}}{\partial x^0}=\left\{h_{ab},H[N,N^a,\pi^{ab},h_{ab}]\right\}, \ \ \ \frac{{\partial \pi}^{ab}}{\partial x^0}=\left\{\pi^{ab},H[N,N^a,\pi^{ab},h_{ab}]\right\},
\ee
which determine the developement in $x^0$ of the fields $h_{ab}$  and $\pi^{ab}$
once $N(x^0,\vec x)$ and $N^a(x^0, \vec x)$ are provided. 

The fields $N(x^0,\vec x)$ and $N^a(x^0,\vec x)$ are not restricted in any way by the field equations. Two different choices of $N(x^0,\vec x)$ and $N^a(x^0,\vec x)$ will generally lead to different functions $h_{ab}(x^0,\vec x)$ and $\pi^{ab}(x^0,\vec x)$, from which apparently different spacetime metric components $g_{\mu\nu}$ can be reconstructed. These two solutions of Einstein's equations can be shown to be related by a diffeomorphism in four dimensions, thereby establishing the connection between the arbitrariness in the choice of $N(x^0,\vec x)$ and $N^a(x^0,\vec x)$ and the freedom in coordinate labelling and the existence of a unique spacetime-geometry $g_{ab}$.

Coordinate-independent information---the true physical content of the formalism---is encoded in quantities that do not depend on the choice of $N(x^0,\vec x)$ and $N^a(x^0,\vec x)$. In this sense, the physical dynamical content of the theory is already contained in the data $h_{ab}$ and $\pi^{ab}$, which solve equation \eqref{hamihami} in a frozen-time manner on the three-dimensional Cauchy slice (the lapse- and shift-dependent spacetime development is redundant). Moreover, different initial data that can be connected by the Hamiltonian flow produced by equation \eqref{evoevo} must be regarded as physically equivalent: the Hamiltonian flow defines gauge transformations \cite{Dirac:1964:LQM}. This statement can be made precise by requiring that any observable quantity $O(h_{ab},\pi^{ab})$ be diffeomorphism invariant, i.e., independent of the spacetime development coded in $N$ and $N^a$. This is achieved by demanding that 
\be\label{cocolico} \left\{O(h_{ab}, \pi^{ab}),H[N,N^a,\pi^{ab},h_{ab}]\right\}=0, \ee
for all $N(x^0,\vec x)$ and $N^a(x^0,\vec x)$. 
In this sense, any observable quantity in a generally covariant theory is encoded in variables that do not evolve with respect to coordinate time, that is, in constants of motion. 
This feature, illustrated here at the classical level,  is sometimes referred to as {\em the problem of time}.

A way to code dynamical information in terms of Dirac observables was proposed via the notion of 
evolving constants of motion, investigated by Rovelli in \cite{Ashtekar1991-ASHCPO, PhysRevD.43.442, Rovelli:1990jm}, and the related idea of partial observables \cite{Rovelli:2001bz, Dittrich:2004cb, Dittrich:2005kc}. The framework we propose here is, in a way, a path integral perspective inspiration of this earlier work.

The reason why {\em the problem of time} is usually ignored in the classical context is that the spacetime development of the initial data provides a practical handle for extracting dynamical information without having to express it in terms of constants of motion \eqref{cocolico}. In other words, the $(3+1)$-dimensional representation provided by \eqref{evoevo} makes it easier, in practice, to identify such diffeomorphism-invariant observables. This is due to the existence of powerful idealizations that allow for the construction of diffeomorphism-invariant (physical) notions point by point in the phase space of gravity, without the need to explicitly solve equation \eqref{cocolico} for generic phase-space functions.

More concretely, once a particular solution of Einstein’s equations is given, physical observables can be extracted, for example, by invoking the idealization of test observers probing the spacetime geometry, or by imposing idealized boundary conditions such as asymptotic flatness, together with a natural collection of `inertial observers' located far away. In applying such idealizations, the spacetime representation becomes crucial and therefore requires solving the evolution equations \eqref{evoevo}.
\begin{figure}[h] \centerline{\hspace{0.5cm} 
\(\begin{array}{c}
\includegraphics[height=5cm]{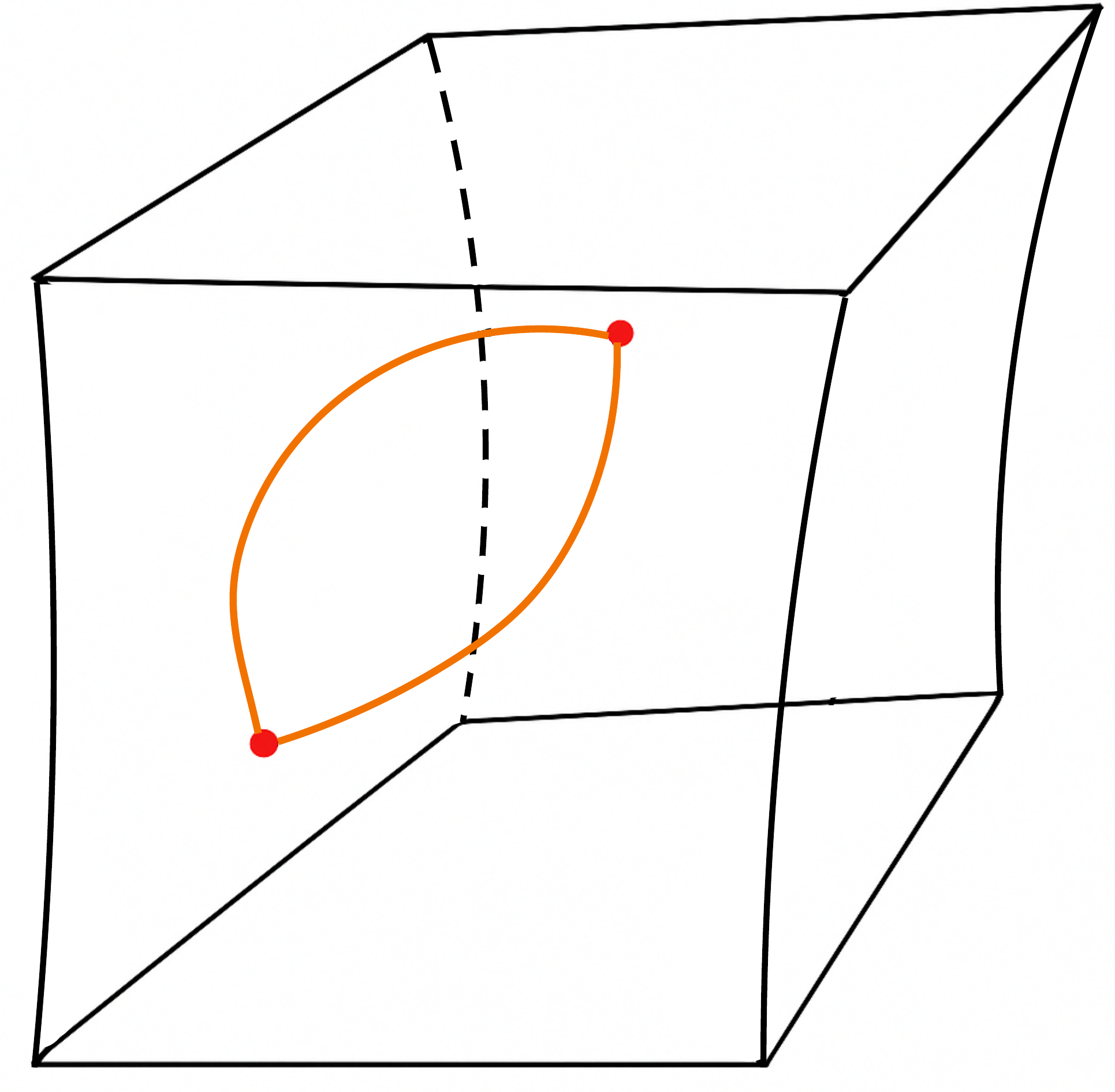} \end{array}
\ \ \ \ \ \ \  \begin{array}{c}
{\rm Diffeo}\\ \longleftrightarrow \end{array} \ \ \ \ \ \ \  \begin{array}{c}
\includegraphics[height=5cm]{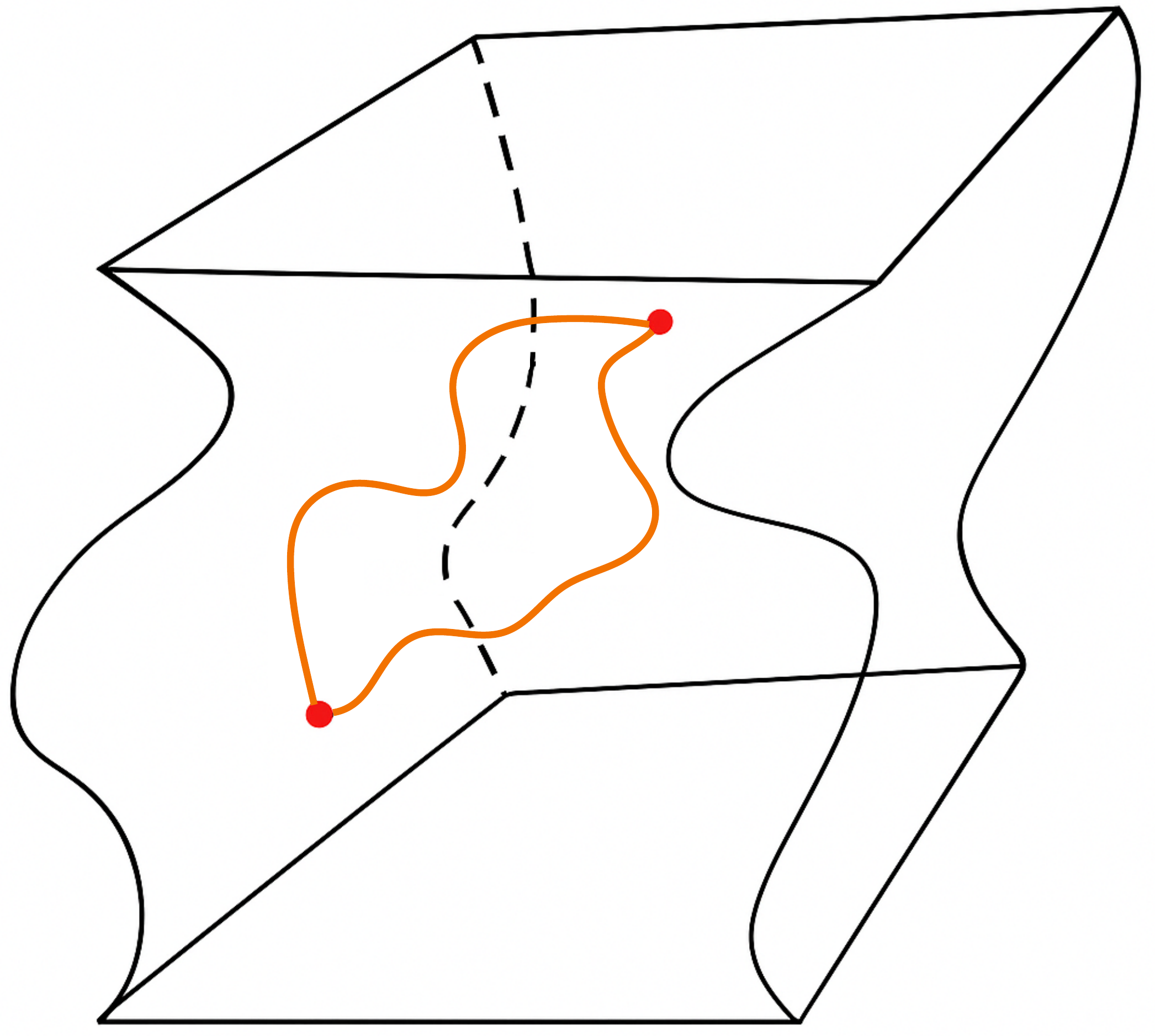} \end{array}\)}
\caption{Same data give rise to different spacetime developments depending on the arbitrary fields $N(x)$ and $N^i(x)$. The two spacetime developements are solutions of Einstein's equations related by a diffeomorphism. }
\label{Tblem}
\end{figure}

A concrete example is illustrated in Figure \ref{Tblem}, where two freely falling test observers are depicted. The observers coincide at an initial event, separate from each other, and then coincide again at a later event. The spacetime representation of this situation depends on the choice of $N(x^0,\vec x)$ and $N^a(x^0,\vec x)$ used in solving \eqref{evoevo}; however, different choices are related by a four-dimensional diffeomorphism. The proper time (i.e., the physical length) between the two coincidence events experienced by either observer is a genuine physical observable: it is diffeomorphism invariant and therefore independent of the spacetime representation of the solution—hence independent of $N(x^0,\vec x)$ and $N^a(x^0,\vec x)$. Even though the proper time along one of the worldline segments is indeed a `constant of motion', namely an intrinsic property of the solution already encoded in the data satisfying \eqref{hamihami}, such a simple physical observable is extremely difficult to represent explicitly as a Dirac observable $O(h_{ab},\pi^{ab})$ satisfying \eqref{cocolico}. This difficulty is precisely why the spacetime representation in the classical theory plays a crucial role in facilitating the physical understanding of `time evolution'.

However, this option is no longer available in the quantum theory for two main reasons.

Physically, the notion of a test observer ceases to be meaningful in regimes where quantum gravity effects become relevant: on the one hand the uncertainty principle precludes the existence of well-defined trajectories, on the other hand any physical degree of freedom gravitates in the strong quantum gravity regime and cannot be considered as a test degree of freedom. Moreover, asymptotic idealizations such as asymptotic flatness have limited applicability: key regions where quantum gravity effects are significant can be causally disconnected from asymptotic infinity (as in the vicinity of black hole singularities), and even the boundary conditions defining asymptotic flatness may break down in the far future in quantum-gravity scenarios—such as black hole evaporation—due to unavoidable departures from semiclassicality \cite{Page:1979tc, Perez:2025tvg}.

Mathematically, the canonical quantization approach exhibits the difficulty clearly.  The quantization rules require one to represent phase-space variables as operators acting on a suitable Hilbert space, where the `dynamics' is defined by the analogue of a time-independent Schrödinger equation of the form
\be\label{qconst}
\widehat H[N,N^a,\widehat\pi^{ab}, \widehat h_{ab}]\ket{\psi}=0,
\ee
for all choices of lapse $N$ and shift $N^a$ (the quantum version of \eqref{hamihami}). Physical states (those for which the previous equation holds) are constant Hilbert space vectors in a quantum theory defined without spacetime.

The flavour of a spacetime representation seems to survive in the path-integral formulation, which can be defined in terms of physical transition amplitudes constructed by averaging over all possible choices of lapse and shift in a pure-gauge time evolution generated by the quantum Hamiltonian. Formally, one writes the physical transition amplitudes as
\ba\label{pi}
\braket{\phi|\psi}_{\rm phys}&\equiv&\braket{\phi|  \int \sD N \sD N^a \exp{\left(-\frac{i}{\ell_p^2} \int \widehat H[N, N^a, \widehat \pi^{ab} ,\widehat h_{ab}] dx^4 \right)}| \psi}\\ \n 
&=& \int \sD h_{ab}\sD\pi^{ab} \sD N \sD N^a \exp{\left(\frac{i}{\ell_p^2}S[N, N^a,h_{ab}, \pi^{ab}]\right)}, 
\ea
where, in the last line, the information about the boundary states is to be encoded in the boundary conditions of the functional integral.
However, even when the previous expression involves summing over geometric degrees of freedom, it is unclear what interpretation one should 
actually give to the histories involved as they represent the averaging over pure gauge degrees of freedom that project the boundary states into the kernel of the quantum constraint \eqref{qconst}.

The above expressions are formal, and serious difficulties arise when attempting to place this prescription on a solid mathematical footing. Some of these difficulties are associated with representing the basic variables on a Hilbert space in the absence of background geometric structures, quantizing the Hamiltonian while preserving the constraint algebra (which stems from the underlying diffeomorphism gauge symmetry) in an unambiguous manner, and incorporating matter fields consistently. Nevertheless, partial progress toward this ambitious program has been achieved within the framework of loop quantum gravity \cite{rovelli_2004, Thiemann:2007zz, Ashtekar:2004eh, Perez:2004hj} and its path-integral representation \cite{Perez:2012wv, Perez:2003vx}, although significant challenges remain open \footnote{There has been remarkable progress in the loop quantization of the gravitational constraints in an anomaly free manner (i.e. respecting the quantum constraint algebra) \cite{Varadarajan:2019wpu, Ashtekar:2020xll, Thiemann:2021hpa, Varadarajan:2022dgg}. Unfortunately, ambiguities of the type discussed in \cite{Perez:2005fn} remain. At present no clear perspective that could help reducing this large ambiguity is well understood.}.

We will not be concerned with such difficult issues here, and will instead simply use the formal structure to illustrate our point. We will see that even when the fundamental quantum-gravitational dynamics is timeless, in the strict sense that all physical observables must be viewed as constants of motion---or quantum numbers labelling quantum states with no apparent spacetime interpretation---spacetime can nevertheless emerge in a suitable limit. This occurs provided that appropriate degrees of freedom are placed in suitable semiclassical states. The results and analysis resonate with those of Rovelli \cite{Rovelli:1990jm} where time (for a generally covariant system) is shown to emerge in a suitable semiclassical state. Our analysis is consistent with that early treatment but it provides new insights. On the one hand it is complementary, as we use the path integral formulation instead of the purely Hamiltonian perspective, on the other hand, time emerges in our formulation from the requirement that the clock degree of freedom be in a semiclassical state while allowing for the rest of the degrees of freedom to remain quantum.  The conclusion, which is largely consensual among those familiar with the Dirac program for quantum gravity, is that no spacetime picture is valid at the fundamental level; spacetime is only an emergent notion under appropriate conditions.

We will illustrate this point using an ultra-simple toy model. Despite its simplicity, the structure of the model is conceptually analogous to that of gravity, and the transparency of the result paves the way for what we believe to be useful conceptual insight into how the spacetime physics we experience around us can emerge, and why it is allowed to take the form it does. Ultimately, this behavior can be traced back to special conditions on the quantum state of the universe, and may even be linked to primordial cosmological questions.

An additional technical value of the results we discuss here is the clarification of an issue that has led to some confusion in the context of spin foams: the cosine problem. We will see that the emergence of time in the semiclassical sector of a clock degree of freedom 
naturally selects the forward propagating amplitude in the path integral of a generally covariant system. This will be explicit in our simple quantum mechanical example. The conclusions can be extrapolated at the formal level to the path integral in quantum gravity and spin foams. 

\section{Approaches to the problem of time}\label{333}

In this section we present a rather brief and partial review of a subject 
with a long history and a very large body of literature which we will most certainly not fairly 
describe here. There are two papers with a large number of references that we refer the reader to
for a more throughout account. One is the classic critical review paper by Kuchar \cite{Kuchar:1991qf}, while the other is the rather recent  paper
by Höhn, Smith, and Lock \cite{Hoehn:2019fsy}, which contains many references and original results that we will briefly comment below.

\subsection{Re-parametrizing: choosing a classical clock}\label{gaugefixing}

One idea to deal with the issue is to introduce a classical clock variable and use it to break general covariance singling out an associated 
Hamiltonian evolution. This is used in approaches to quantum gravity; however, it is convenient to illustrate the idea
first in a much simpler context capturing the basic features relevant in the construction: the relativistic particle.
The relativistic particle action is analogous to general relativity in  one dimension \cite{BRINK1976471}, and 
therefore shares  with the gravitational theory the issue of time (in a suitable sense). This is clear from the reparametrization invariance of 
the relativistic particle action
\be\label{rela}
S=-m\int\sqrt{|g^{(1)}|} d\lambda=-m\int d\lambda \sqrt{-\eta_{\mu\nu} \dot x^\mu\dot x^\nu}, 
\ee
where $g^{(1)}$ is the induced metric along the world-line of the particle. The associated Hamiltonian is identically vanishing. Indeed using $p_\mu\equiv \partial L/\partial \dot x^\mu=-m \eta_{\mu\nu} \dot x^\nu/ \sqrt{-\eta_{\alpha\beta} \dot x^\alpha\dot x^\beta}$ one finds that  $H\equiv p_\mu\dot x^\mu- L\approx 0$ or equivalently 
\ba\label{shelly}
C\equiv p_\mu p^\mu+m^2\approx 0.
\ea
Thus, the phase space formulation of the action is
\be
S=\int  d\lambda \left[p_\mu\dot x^\mu-N(p_\mu p^\mu+m^2)\right],
\ee
with $N(\lambda)$ a Lagrange multiplier which is the strict analog of the shift 
field of its higher dimensional gravitational relative. The theory has no notion of time evolution 
and suffers of the same issues as general relativity. However, it is very natural to think of the relativistic particle, not as a truly isolated system, but as one evolving within the laboratory time frame, where we can use a special parametrization where $
\lambda=x^0=t$. Such `gauge fixing' can be formally thought of as the one that introduces the suplementary condition $\chi\equiv x^0-\lambda=0$ that renders the new constraint system $C=0$ and $\chi=0$ second class. One solves the constraints and finds that $p_0=\pm \sqrt{\vec p\cdot \vec p+m^2}$ from which the positive energy branch can be written as
\be\label{Sch1}
p_0+\sqrt{\vec p\cdot \vec p+m^2}=0.
\ee
After substitution in the previous phase space action  we obtain (for the $-$ solution branch) 
\be
S=\int dt \left(\vec p \cdot \dot{\vec x}-\sqrt{\vec p\cdot \vec p+m^2}\right),
\ee
with Hamiltonian $H=\sqrt{\vec p\cdot \vec p+m^2}$ defining laboratory time evolution. 
Indeed, after quantization, either formally viewing \eqref{Sch1}  as a constraint with $p_0=-i\hbar\partial/ \partial t$ (or canonically quantizing the previous action) one obtains the Schrödinger equation in laboratory time for the wave function of the particle.
The fact that there are two branches is related to the well known `negative energy' problem of relativistic 
quantum mechanics, which can only be consistently solved by renouncing the quantization of a single relativistic particle and considering instead the quantum theory of fields \cite{BjorkenDrell1965} (a first indication that the notion of particle is not well defined in the relativistic context \cite{Wald:1995yp}).


The previous is a cartoon illustration of the idea of using some degrees of freedom as a clock variable to resolve the problem of time in general relativity as proposed in \cite{Brown:1994py},  where the clock degrees of freedom are assumed to be given by a dust fluid. This perspective has been implemented using scalar fields in the context of quantum cosmology, in loop quantum cosmology \cite{Ashtekar:2011ni}, as wells as in full  quantum gravity in \cite{Giesel:2007wn, Domagala:2010bm, Giesel:2016gxq}. Here we follow the particularly clear notation of the last references to review it.
In a few lines, one observes that for general relativity coupled to a scalar field the diffeomorphism constraint can be written as
\be
C_a(x)=p_\phi \nabla_a\phi+C_a^{\rm rest}(x)=0, 
\ee
where $p_\phi$ is the momentum conjugate to the scalar field $\phi$, and $C_a^{\rm rest}$ is the contribution to the scalar constraint from all the other degrees of freedom (including gravity).  Similarly, calling $C_{\rm rest}(x)$ the contributions other than the scalar field, the scalar constraint is
\ba\label{pipina}
\sqrt{h} C(x)&=& \frac{p_\phi^2}{2}+\frac{h}{2}\, h^{ab}\nabla_a\phi\nabla_b\phi+h\, V[\phi]+\sqrt{h} C_{\rm rest}(x)\\
\n &=&\frac{p_\phi^2}{2}+\frac{h\, h^{ab}}{2 {p_\phi^2}}(C_a(x)-C^{\rm rest}_a(x))(C_b(x)-C^{\rm rest}_b(x))+h\, V[\phi]+\sqrt{h} C_{\rm rest}(x)=0,
\ea
where in the second line one has used the vector constraint to solve for $\nabla_a\phi$. After solving for $p_\phi$, they propose the alternative scalar constraint
\be\label{ete}
p_\phi+H=0,
\ee
with a `true Hamiltonian'
\be
H=\sqrt{-\sqrt{h} C_{\rm rest}+ \sqrt{h}\sqrt{ C_{\rm rest}^2-h^{ab} C^{\rm rest}_a C^{\rm rest}_b} },
\ee
that defines `time' evolution in scalar field time. In other words, in the functional representation of the new quantum constraint \eqref{ete}---with $p_\phi=-i\hbar \delta/\delta \phi$---one recovers a Schrödinger equation as in \eqref{Sch1}. Leaving aside the issue related to the other three branches of solutions of \eqref{pipina} (which is a quartic equation for $p_\phi$), the most important weakness of this proposal is the fact that the `time' variable---identified with the scalar field $\phi(x)$---is a dynamical degree of freedom that is never quantized.
As we will see, as a result of our analysis, in simple systems one can hope that such a prescription would correspond to a suitable semiclassical corner of the theory, solely emerging if the clock variable happens to be in a very special clock state. However, quantum fluctuations in the clock degree of freedom  (expected to be large in full quantum gravity regimes, e.g. near singularities) have physical consequences that are completely erased by the present construction.  Different clock choices will lead to different quantum dynamics: different Hamiltonians. While this is physically acceptable at the classical level—with the timeless classical theory providing the unifying structure relating alternative dynamics—it is unclear how this can be understood at the quantum level if one renounces the quantization of the fundamental timeless theory. For these reasons, it is also unclear how the present strategy can lead to trustworthy predictions at the fundamental scale, at least when one has no access to the fundamental timeless quantization. Under suitable circumstances, however, the time-neutral quantum theory plays a central role in the interpretation of its parametrized offspring \cite{Hohn:2018toe}.

\subsection{Defining a fundamental quantum physical clock}\label{Pauli}

In the previous section we reviewed a proposal for the resolution of the problem of time in generally covariant systems, such as the relativistic particle (gravity in 1d) or general relativity in arbitrary dimensions, based on the idea of choosing a clock variable and describing the dynamics relationally with respect to this degree of freedom at the classical level. The procedure is performed as a first step towards the quantization of the resulting parametrized system. While this perspective may be sensible in the classical context—where, as argued in the introduction, it is not really needed—it suffers from the important drawback that the chosen time variable remains classical upon quantization. This is a serious weakness in view of constructing a consistent theory of quantum gravity, where all degrees of freedom must be quantum at the fundamental level.

Trying to avoid such difficulties, one often takes the opposite view starting from the onset with a purely quantum notion of ideal clock. Formal discussions of such quantum mechanical notion of time make the assumption of the existence of a time self adjoint operator $\widehat T$ associated to a clock degree of freedom with selfadjoint Hamiltonian $\widehat H_c$ and commutation relations $[\widehat T, \widehat H_c]=i\hbar$. If such a quantum degree of freedom were available, with the constraint of the system describing the clock and the rest of the universe becoming
\be\label{otro}
H_c+H_{\rm rest}=0,
\ee
then the constraint would formally look like the Schrödinger equation.  
The key difficulty with \eqref{otro} is that no realistic degree of freedom can satisfy the previous requirements for an ideal quantum clock. For instance, self adjointness of $\widehat T$ implies that if $\widehat H_c\ket{E_c}=E_c\ket{E_c}$, then $\ket{E_c+\alpha} = \exp{i\alpha \widehat T} \ket{E_c}$, and thus the spectrum of $\widehat H_c$ cannot be bounded below---as in any realistic physical model of a clock.   In their seminal paper,  Page and Wootters \cite{PW} introduced an alternative avenue to recover dynamics  by defining conditional probabilities that encode time evolution relationally. Specifically, they asked: what is the probability of a given degree of freedom to take a value when another one (seen as a clock) takes some other value. 

For the sake of the argument that will be developed below, there are two aspects that we would like to emphasize here. 

First, note that one could have instead started directly with a realistic Hamiltonian that is bounded below.  Perhaps the simplest choice would be a free particle with $ H_c= p_c^2/(2m)$. Then one would try at defining a `time' variable using  $T=m  p_c^{-1}  x_c$, which satisfies $\{T, H_c\}=1$. The loophole with respect to the previous argument is that the proposed time variable (aside from factor-ordering ambiguities)  cannot be represented as a self-adjoint operator because of the singular behaviour of $p_c^{-1}$. Nevertheless, a semiclassical notion defined in terms of mean values, namely $t_{\rm phys}\equiv m \braket{x_c}/\braket{p_c}$, in some suitable state still makes sense as long as the clock state  satisfies $\braket{p_c}\neq 0$. Moreover, for suitable states such a time variable may approach the notion introduced in Section \ref{gaugefixing}, though with quantum fluctuations which cannot be neglected in general. 

Second, note that the time operator does not commute with the total Hamiltonian (nor the variables used in the mean field definition of the previous paragraph). More precisely, the observation of these variables does not preserve the solution space of the Hamiltonian constrain \eqref{otro}: they are not Dirac observables. This appears as a serious issue as their measurement  would physically require an external intervention violating the assumed closed-ness of the system (imposed to us in generally covariant systems by diffeomorphism invariance). In \cite{Gambini:2008ke} this problem is analyzed by constructing the Page and Wootters formalism
 in terms of the notion of evolving constants of motion which are labelled by partial observables and defined first classically via the resolution of the classical evolution equations. 
We will see in Section \ref{ddd} that our proposal provides a different perspective on this issue where dynamics is coded directly at the quantum level in terms of Dirac observables without having to resolve the classical equations of motion.

In reference \cite{Hoehn:2019fsy}, some of the issues mentioned above are dealt with by extending the concept time variable in the quantum theory  to the more general notion of  positive operator-valued measures (POVM). This allows for the inclusion of physically realistic bounded below clock Hamiltonians $H_c$.  The second issue raised above is also shown to be resolved in their formulation, as the Page-Wootter-like conditional probabilities  defined via the POVM are shown to correspond to gauge invariant relational observables (even when the interpretation of these remains dependent on apparently kinematical concepts). These observables 
are Dirac observables which in the situations at hand are labelled by data that are not coded in terms of gauge invariant notions but rather partial observables as put forward in \cite{Rovelli:2001bz}. Finally, the emphasis on probabilities in the POVM approach---whose meaning in a closed universe brings us back to the hard problem we have decided not to get engaged with in this paper---is to be remarked as well.
This paper is certainly central for the discussion we are focusing on, in particular it would appear to provide a general constructive framework in contrast with our example-based type of analysis (which is certainly one of its limitations). Nevertheless, the picture we propose differs from the later in that we do not invoke any notion of time operator or POVM but rather find internal time as a semiclassical notion arising only in suitable situations dynamically (in computation of physical amplitudes with the path integral). The advantage we see is the fact that the formulation is well suited for the analysis of the functional integral approach which is one of the main dynamical approaches in non perturbative treatments of quantum gravity.  

In our view some aspects of the situation can become more intuitive by taking the third path just mentioned: the one that lies between the completely classical view of Section \ref{gaugefixing} and the fully quantum perspective of the present section. We will argue for the advantages of such an approach by illustrating it in the context of a manageable nonrelativistic toy model. Even though in the model we will make some concrete choices for the sake of clarity, the argument will remain conceptually general and, in principle,  extendable to the generally covariant setting of quantum gravity---provided that the theory is formulated along the usual lines of Hilbert spaces, states, and projections onto the kernel of quantum constraints.

We will see that this mean-field notion of time is the one that naturally emerges when describing the projection of kinematical states onto the kernel of the quantum constraints via the path-integral representation of an otherwise timeless system, as long as the relevant conditions on the state of some clock degree of freedom are satisfied. We will see that the time variable is genuinely an internal time which can be precisely coded in suitable Dirac observables. At the fundamental level and for generic states, there will be no meaningful notion of time, and physics will have to be understood in a timeless manner. This is not a problem; it is simply the way quantum generally covariant systems behave.

\subsection{A third path: Emergent semiclassical time from the path integral representation (a non relativistic realization)}\label{pipi}

The path integral representation offers an insightful perspective because it allows one to address aspects of the problem of time without directly confronting the measurement problem and the probabilistic interpretation of quantum mechanics, which are nontrivial issues that cannot be ignored in discussions of the quantum physics of a closed system (or of the universe as a whole in the case of quantum gravity). The path integral viewpoint is helpful in circumventing, to a certain extent, these difficulties, in a way similar to how it provides insight into the classical limit of a quantum theory without resolving all of the underlying conceptual issues. Interpretational questions in quantum mechanics for a closed universe are important and will, unfortunately, remain beyond the scope of the present analysis.

\subsection{The problem of time in a closed system: physical discussion}

A non-relativistic closed system with fixed energy $E$ is the simplest analogy of the situation in general relativity. 
The quantum system is assumed to be represented by those states $\ket{\psi} \in \sH$ satisfying the 
constraint equation 
\begin{equation}\label{concon}
C\ket{\psi} \equiv ({H} - E\, \mathbb{I} )\ket{\psi} = 0,
\end{equation}
for a given eigenvalue $E$ of the Hamiltonian $H$. 
For the moment we assume that the eigenvalue $E$ is in the discrete part of the spectrum of the Hamiltonian so that states satisfying \eqref{concon} are normalizable. An orthonormal basis of solutions of the previous equation---the time-independent Schrödinger equation with a fixed energy eigenvalue $E$---is denoted $\ket{E,n}$, where $n$ represents an additional quantum number labelling the degeneracy of the eigenvalue $E$ (which can always be taken as an integer assuming separability of the Hilbert space). The quantum number $n$ can correspond to the rearrangement of a large number of quantum numbers associated to a complete set of commuting observables (which correspond to the number of degrees of freedom in the closed system). Such number of degrees of freedom can be very large.
The most simple statement of the problem of time is the statement that any density matrix defining the state of the closed system is time independent.  Namely, 
\be
\rho=\sum_{n,m} \rho_{nm} \ket{E,n}\bra{E,m},
\ee
for which $[\rho, H]=0$  for  any arbitrary choice of $\rho_{nm}$. Time evolution as generated by $H$ acts as a global phase in the states of the closed system and becomes simply unobservable. 

One can restate the previous simple physical fact into the language of gauge theories used in the statement of the problem of time in the gravitational context as follows: the Hilbert space of solutions of equation \eqref{concon},  $\sH_{\rm phys}\subset \sH$, is called the physical Hilbert space of the closed system while the full Hilbert space $\sH$ is called the Kinematical Hilbert space. With this notation the previous  orthonormal basis of states $\ket{E,n}$ is a basis of  $\sH_{\rm phys}$.  One can define the so-called physical inner product  between arbitrary states in $\sH$ as follows
\be\label{pip}
\braket{\alpha|\beta}_{\rm phys}\equiv\braket{\alpha|\delta_{H,E}|\beta},
\ee
where $\delta_{ H,E}$---defined as
\be
\delta_{\widehat H,E}=\delta_{H,E}^\dagger=\sum_n\ket{E,n}\bra{E,n},
\ee
with $\delta_{\widehat H,E}^2=\delta_{\widehat H,E}$---is the projector from the kinematical Hilbert space $\sH$ to the physical
Hilbert space of the closed system $\sH_{\rm phys}$. The physical inner product defines the probability amplitudes of the closed system. 

The constraint \eqref{concon} is the strict analog of the Hamiltonian constraint of general relativity. 
The gauge symmetry associated to the action of the previous constraint corresponds to the rescaling of the quantum states $\ket{\psi}\in \sH_{\rm phys}$ by the action 
\be
\exp(-iN H)\ket{\psi}=\exp(-i N E)\ket{\psi},
\ee
which (as already emphasized) has no observable consequence as it corresponds to a global constant phase. 
Observables $O$ of the closed system are those operators that commute with the Hamiltonian $H$, as these are the only whose observation does not change the value of the system's energy (only energy conserving observations within the closed universe are doable and physically relevant). Such physical restriction coincides in this simple model with the mathematical demand that observables be gauge invariant,
namely,
\be
\exp(iN  H) O\exp(-iN  H)= O,
\ee
or equivalently
\be
[ O, ({H} - E\, \mathbb{I} )]=0,
\ee 
which defines Dirac observables in the usual mathematical sense of gauge theories.
 It is often (rightly) emphasized that energy is a subtle notion not always well defined in general relativity. The Hamiltonian formulation of gravity 
provides a trivial answer to what energy is: it is identically zero in background independent field theories; the ultimate 
notion of closed systems.  From the physical perspective, the condition that Dirac observables commute with the Hamiltonian \eqref{hamihami}---independently of the choice of lapse $N$ and shift $N^a$---is (the multi-fingered) version of `energy conservation' in a closed system,
whose quantum version is \eqref{concon}.

\subsection{The path integral definition of physical amplitudes for a closed system}

Now we construct the path integral representation of the physical inner product \eqref{pip}. 
For a non relativistic system one can write the action principle in a reparametrization invariant manner 
as follows
\be
S[x_i]=\int L\left(x_i,\frac{dx_i}{dt}\right) dt=\int L\left(x_i,\frac{\dot x_i}{\dot t}\right) \dot t  d\lambda,
\ee
where $\cdot$ denotes derivative with respect to an arbitrary parameter $\lambda$
\footnote{Note that this is what we would get in the low velocity limit of the free relativistic particle action \eqref{rela}  written in inertial coordinates where we call $x^0=t$, namely 
\be
S=-m \int \dot t \sqrt{ 1-\frac{\dot {\vec x}\cdot \dot {\vec x}}{\dot t^2}} d\lambda.
\ee }. Hamiltonian analysis of the previous action implies that the phase space variables---extended to include the conjugate pair $(t\equiv x^0, p_0\equiv \partial L/\partial \dot t ))$---satisfy the constraint
\be\label{constraint}
p_0+ H(x_i,p_i)=0,
\ee 
where $H(x_i,p_i)$ is the standard Hamiltonian in the original formulation. The corresponding action principle in the Hamiltonian form is
\be
S[N,x^0,p_0, x_i,p_i]=\int d\lambda \left(-\dot p_0  x_0 +\sum_i p_i \dot x_i- N\left(p_0+H(x_i, p_i)\right)\right),
\ee
with $p_0$  a conserved quantity if the original Lagrangian  is time independent (as the assumption of closedness implies).  Note also that we have written the canonical term $-\dot p_0  x_0$ with the derivative acting on $p_0$ only for later convenience (the alternative option $p_0  \dot x_0$ differing from ours by a total derivative not affecting the dynamics). As in gravity, the Lagrange multiplier $N(\lambda)$ is not determined by the equations of motion, this is due to the gauge symmetry associated to reparametrization invariance which is generated by the Hamiltonian constraint \eqref{constraint} that, here,  follows from $\delta S/\delta N=0$. 

In order to write the path integral definition of the quantum propagator one needs to introduce a restriction on the Lagrange multiplier in order to avoid trivial divergencies due to the redundant imposition of the previous constraint (which is a constant of motion) at different values of $\lambda$. Reparametrization invariance of the action implies that under $\lambda\to f(\lambda)$ the Lagrange multiplier $N\to N /\dot f$. The simplest choice is to impose the condition $\dot N=0$ which completely fixes the freedom with the boundary condition that the initial and final values of $\lambda$ remain fixed \cite{Teitelboim:1981ua}. 
The quantum theory can now be defined via the path integral formalism as follows
\ba\label{16}
&& \int dN \braket{p^{\rm\va  out}_0 x^{\rm out}_i|  \exp{\left(-\frac{iN }{\hbar}(\widehat p_0+\widehat H(x_i, p_i))\right)}|p_0^{\rm in}, x^{\rm in}_i} = \\ && \ \ \ \ =\int dN \sD p_0\sD x_0 \sD p_i \sD x_i \exp{\frac{i}{\hbar}\int d\lambda \left(-\dot p_0  x_0 +\sum_i p_i \dot x_i- N\left(p_0+H(x_i, p_i)\right)\right)},\n 
\ea 
where the integration over the (constant) Lagrange multiplier $N$ imposes the quantum version of the constraint  \eqref{constraint}.

Note that it is essential to integrate the variable $N\in \R$ over its full range in order for the amplitude to correctly impose the Hamiltonian constraint \eqref{constraint}. A modified amplitude, obtained by restricting the integration of the lapse function to the range $N\in \R^+$, is argued in \cite{Teitelboim:1981ua} to provide the proper physical definition. This argument is based on the observation that such a restriction yields the Feynman propagator in the case of the free relativistic particle action (of which ours corresponds to the low-velocity limit in the absence of an external potential).  The Feynman propagator, being a Green function of the Klein-Gordon equation, plays a central role in defining perturbation theory in the presence of interactions in quantum field theory. In the language of Feynman diagrams,  off-shell contributions of the Feynman propagator (not satisfying \eqref{shelly}) associated with internal lines are necessary in computing physical processes in interacting QFT. In the quantum theory of a single relativistic particle, however, and according to the standard Dirac quantization procedure, gauge invariance must be imposed strongly by defining the physical Hilbert space satisfying \eqref{shelly}. This requirement is implemented through integration over the full range of the lapse function $N$.  A similar restriction to positive lapses in the gravitational path integral formulation was imposed 
in  \cite{Teitelboim:1981ua}  with the aim of constructing an exotic definition of quantum gravity using perturbation theory in the signature around `zero signature gravity'. Such restriction is incorrect in view of defining a non perturbative quantum path integral.

The author of \cite{Teitelboim:1981ua} was certainly aware of the previous.  However, he was interested in developing a suitable perturbation theory of quantum gravity---where the notion of Feynman propagator would rightly play a key role---based on the signature as perturbation parameter. Such perturbative approach has remained formal ever since; however, the idea of restricting histories of the gravitational field to positive definite lapses in order to incorporate  somehow a necessary `causal' order has survived \cite{Livine:2002rh, Engle:2011un, Engle:2012yg, Vojinovic:2013faa, Engle:2015mra, Immirzi:2016nnz,  Bianchi:2021ric, Jercher:2022mky}. We will discuss the implications of our analysis in greater detail in Section \ref{coseno}. We will argue that the notion of forward propagation, which has often been sought by modifying the definition of transition amplitudes, is consistently recovered without modifying the fundamental definition of the transition amplitudes when boundary quantum states are suitable. 

In order to make contact with the previous section, we can define the quantum theory of a closed system by restricting the energy to a fixed value $E$. In the present situation this corresponds to assuming that the constant of motion $p_0$  equals $-E$, or equivalently (at the quantum level) that the wave function corresponding to the configuration variable $x^0$ is an eigenstate of $p_0$, namely
\be
\widehat p_0\ket{E, \cdots}=-E \ket{E,\cdots}, 
\ee 
This amounts to setting the boundary values $p_0^{\rm in}=p_0^{\rm out}=-E$ in transition amplitudes. Thus, for a closed system with fixed total energy $E$ equation \eqref{16} yields \ba
\label{19}&& \!\!\!\!\!\!\!\!\! \braket{x^{\rm out}_i| x^{\rm in}_i}_{\rm phys.} \equiv  \int \frac{dN} {t_p} \braket{E, x^{\rm out}_i|  \exp{\left(-\frac{iN}{\hbar}(H(x_i, p_i)-E)\right)}|E, x^{\rm in}_i} =\frac{2\pi \hbar}{t_p}\braket{E, x^{\rm out}_i|  \delta(H(x_i, p_i)-E)|E, x^{\rm in}_i}
\n  \\
&\equiv& \int \frac{dN} {t_p} \int \sD p_i \sD x_i \exp{\frac{i}{\hbar}\int d\lambda \left(\sum_i p_i \dot x_i- N\left(H(x_i, p_i)-E\right)\right)},
\ea 
where we recover the definition of the physical inner product \eqref{pip}, and its path integral representation in the last line. 
The time scale $t_p$ in the lapse integration measure---naturally associated with the Planck time---is necessary to get a dimensionless physical amplitude. The energy $E$ 
is now allowed to be in the continuous part of the spectrum of $H$. From now on we 
focus on the closed system with energy $E$---satisfying \eqref{concon}---so we will drop the $E$ label in the kinematical states, namely $\ket{E,x_i}\to \ket{x_i}$. For further use we write
 \be\label{citar}
 \braket{x^{\rm out}_i| x^{\rm in}_i}_{\rm phys.}={\hbar} \int \frac{dN} {t_p} \int \overline{\sD }x_i \exp{\frac{i}{\hbar}\left[\int\limits_0^{N } L\left(x_i,\frac{dx_i}{dt}\right) dt \right]}={\hbar}\int \frac{dN} {t_p}\left(\int \overline{\sD }x_i \exp{\frac{i}{\hbar}S\left[N , x_i(t)\right]}\right), \ee
where we have integrated out the momenta (assuming that the kinetic term is quadratic in $p_i$) in the last line of \eqref{19} in order to obtain the configuration space path integral from the phase space one
with the usual modified configuration space measure ${\sD} x_i\to \overline{\sD} x_i$ (see for instance \cite{kleinert}).

\section{The `cosine problem' } \label{coseno}

The path integral representation of gravity requires the functional integral to include the integration of the Lagrange multiplier fields 
$N$ and $N^a$. We can write, at the formal level, the physical transition amplitude (or physical inner product, analog of \eqref{19}) as
\ba\label{cosy}
    \langle \psi_f \lvert \psi_i \rangle_{\rm phys}
  &=&\int \sD N \sD N^a \sD h_{ab} \sD \pi^{ab} \cos{\left[\int\limits dx^4 \frac{N}{\ell_p^2}  \sqrt{h} \, \left(R^{\va (3)}[h_{ab}]+K_{ab} K^{ab}-K^2\right)\right]}  ,
\ea
where the cosine appears due to the symmetry
\be\label{sysy} S[-N, N^a, h_{ab}]=-S[N, N^a, h_{ab}],\ee 
and the invariance of the lapse integration measure. 
Such symmetry is obvious when using the Einstein-Hilbert action in its form \eqref{action} 
recalling that the extrinsic curvature can be explicitly written as \cite{Wald:1984rg}
\be
K_{ab}=\frac{\dot{h}_{ab} - D_a N_b- D_b N_a}{2 N}, 
\ee
for $D_a$ the covariant derivative compatible with the space metric $h_{ab}$. 
The phase space path integral above coincides with the one in canonical form in \eqref{pi} which uses the action in the form \eqref{caction}. The symmetry invoked in the argument leading to the cosine is of course valid for \eqref{caction} as well, namely $S[-N, N^a,h_{ab}, \pi^{ab}]=-S[N, N^a,h_{ab}, \pi^{ab}]$. Here we used the non canonical form in order to see this more  transparently. 

The formal structure illustrated in the previous paragraph reemerges in the spin foam proposals \cite{Perez:2012wv, Perez:2003vx} that aim at giving a concrete definition of the path integral. The fact that both `forward', $\exp(iS[N,N^a, h_{ab}, \pi^{ab}])/\ell_p$, and `backward', $\exp(-iS[N,N^a, h_{ab}, \pi^{ab}])/\ell_p$, propagating amplitudes contribute on equal footing is sometimes regarded as a problem. 
More concretely, the amplitude for the transitions $\psi_i \to \psi_f$ is given by a sum over all possible spin foams $\mathcal{F}$---Feynman-like higher dimensional graphs consisting of two complexes labelled by geometric quantum numbers---connecting the boundary states. Namely,  
\begin{equation}
    \langle \psi_f \lvert \psi_i \rangle_{\rm phys} = \sum_{\mathcal{F}} \mathcal{A}[\mathcal{F}],
\end{equation}
\noindent where $\mathcal{A}[\mathcal{F}]$ is the weight assigned to each spin foam $\mathcal{F}$, which can be written as a product of basic amplitudes called vertex amplitudes ${\cal A}_{\rm vertex}$, associated with the fundamental building blocks of the underlying two-complex  ($d$-simplices in $d$ dimensions). In this way, the spin foam formalism mirrors the path integral approach by summing over histories coded in combinatorial structures (seen as the counterpart of quantum geometries). Just as in the formal path integral, `time-reversal symmetry'---reinforced by the form of the fundamental theory in a way that is the analog of the sum over positive and negative lapse and the corresponding discrete symmetry of the underlying amplitudes---is preserved unless explicitly broken. 

This is explicit in the  analysis of Ponzano and Regge \cite{Ponzano:1968} of a spin foam model for 3d quantum gravity. More precisely, the semiclassical limit of the vertex amplitude (the quantum amplitude of a single element of discrete spacetime history) is obtained by studying its large spin assymptotic, where a discrete geometric interpretation emerges, and comparison with the discrete Regge model \cite{Regge:1961px} is possible. Schematically, instead of  $\mathcal{A}_{\text{vertex}} \sim e^{iS_{\text{Regge}}}$, the time-reversal invariance (analogous to that present in the continuous action \eqref{sysy}) leads to 
\begin{equation}
    \mathcal{A}_{\text{vertex}} \cong \frac{1}{\sqrt{\lvert \det( H_{S_{\rm Regge}}) \lvert}}\cos \left({S_{\text{Regge}}}+\frac{\pi}{4}\right),
    \label{eq:VertexSF}
\end{equation}
\noindent where $H_{S_{\rm Regge}}$ is the Hessian of the Regge action. This is the way the {\em cosine problem}  arises in spin foam models.
Similar behaviour has been identified in the semiclassical limit of the EPRL model \cite{Engle:2007wy} for 4d quantum gravity \cite{Dona:2020yao} with the phase inside the argument of the cosine taking a non trivial value $-\arg\left[\det H_{S_{\rm Regge}}\right]/2$ (see also \cite{Barrett:2009gg}).

While several proposals aim to resolve the cosine problem by modifying the amplitudes to enforce forward propagation \cite{Livine:2002rh, Engle:2011un, Engle:2012yg, Vojinovic:2013faa, Engle:2015mra, Immirzi:2016nnz, Bianchi:2021ric, Jercher:2022mky}, we take a different standpoint: this feature is not a pathology, but rather an expected outcome of a fundamentally background-independent formulation of quantum gravity amplitudes. The forward propagation that one is used to in the familiar QFT can only emerge under suitable circumstances from the background independent
timeless fundamental theory when transition amplitudes of very particular quantum states are considered.
The standard forward propagation in time of QFT on a given spacetime background (or in quantum mechanics) is, we will argue,  a property stemming from features of the underlying states of quantum gravity from which the non gravitational effective treatment emerges. 

In order to argue this, we first start by showing the transition amplitudes in a non relativistic closed system defined by \eqref{concon} are also
controlled by the cosine of the action describing the system.
More precisely,  assuming the underlying Lagrangian is quadratic (or in general, in the stationary phase approximation), we can write \eqref{citar} as
 \be
 \braket{x^{\rm out}_i| x^{\rm in}_i}_{\rm phys.}={\hbar}\int \frac{dN}{t_p} \left(F[N]  \exp{\frac{i}{\hbar}S\left[N , x^{\va \rm cl}_i(t)\right]}\right), 
\ee
where $S\left[N , x^{\va \rm cl}_i(t)\right]$ is the action evaluated on the solution of the classical equation on motion satisfying the boundary conditions $x_i(0)=x^{in}_i$ and $x_i(N )=x^{out}_i$, and  
\be F[N ]=\sqrt{\left|\det\left(\frac{i}{2\pi \hbar}\frac{\partial^2 S\left[N , x^{\va \rm cl}_i(t)\right]}{\partial x_i^{\rm in}\partial x_j^{\rm out}}\right)\right|} \exp\left(i\frac{\pi}{4} n\, {\rm sign}(N )\right) \ee is the fluctuation factor, where $n$ is the number of degrees of freedom involved. 

As in the gravitational case, time reversal invariance implies that \be S\left[N , x^{\va \rm cl}_i(t)\right]=-S\left[-N , x^{\va \rm cl}_i(t)\right]\ \ \ {\rm and} \ \ \ F[N ]=F[-N  ] \exp\left(in\frac\pi2\right),\ee
and thus
\be\label{coscos}
{\braket{x^{\rm out}_i| x^{\rm in}_i}_{\rm phys.}={\hbar}\int \frac{dN}{t_p} \left(|F[N]|  \cos\left({\frac{1}{\hbar}S\left[N , x^{\va \rm cl}_i(t)\right]}+n\frac{\pi}{4}\right)\right)},
\ee
which relates the physical inner product with the cosine of the action evaluated at the classical solution.
The formal structure of the model is the same as the one of general relativity \eqref{cosy}. The emergence of the forward evolution for suitable boundary states (which will involve suitable states of a clock degree of freedom) will be explicitly seen in what follows.

\section{quantum clocks and  physical time}\label{calculos}

We revisit the previous discussion assuming a degree of freedom in the closed system to be in a
suitable semiclassical state (a good-clock state). We will show how internal dynamics can emerge together with the associated forward
time evolution that resolves the cosine tension.

\subsection{Clock time and the emergence of dynamics in a closed quantum system}

We  idealize the clock system as a non interacting degree of freedom so that the total Hamiltonian can be written as
\be\label{lack}
H=H_{\rm c}+H_{\rm R},
\ee 
where $H_{\rm c}$ stands for the clock Hamiltonian and $H_{\rm R}$ for the hamiltonian of the rest of the degrees of freedom in the closed system. The kinematical Hilbert space is assumed to be the tensor product $\sH=\sH_{\rm c}\otimes\sH_{\rm R}$.  We denote $\psi^{\rm in}\in \sH_{\rm c}$ and $\psi^{\rm out}\in \sH_{\rm c} $ and consider the physical amplitude  (from \eqref{19}) 
%
\ba\label{ac}
\braket{\psi^{\rm out}_{\rm c};x^{\rm out}_i| x^{\rm in}_i; \psi^{\rm in}_{\rm c}}_{\rm phys.} &\equiv& 
\frac{{2\pi \hbar }}{t_p} \braket{\psi^{\rm out}_{\rm c}, x^{\rm out}_i|  \delta(H_{\rm R}(x_i, p_i)+H_c-E)|\psi^{\rm in}_{\rm c}; x^{\rm in}_i} = \\
&=&\int \frac{dN} {t_p}  \underbrace{\braket{\psi^{\rm out}_{\rm c}|\exp{\left(-i \frac{N }{\hbar} H_c \right) }\psi^{\rm in}_{\rm c}}}_{ A_{\rm c}(N)\equiv \rm clock\ amplitude}  \bra{x^{\rm out}_i}\exp{\left(\frac{-iN } {\hbar}(H_{\rm R}(x_i, p_i)-E)\right)} \ket{x^{\rm in}_i}, \n
\ea
The first line establishes the entanglement between the clock degree of freedom and the rest of the universe imposing \eqref{concon}, while in the second line we have introduced the definition of the clock amplitude $A_c(N)$. A key observation of these notes is that for suitable boundary clock states
(what we will refer to as the {\em good-clock limit}), the clock amplitude $A_c(N)$ will be strongly peaked at the values of the clock time-defining degree of freedom 
\be \label{wea}
N \approx \tau_{\rm phys},
\ee
where $\tau_{\rm phys}$ corresponds to what we call physical time.
The previous statement implies that
the physical inner product becomes
\be
{\braket{\psi^{\rm out}_{\rm c};x^{\rm out}_i| x^{\rm in}_i; \psi^{\rm in}_{\rm c}}_{\rm phys} \approx 
\exp{\left(-i\frac{\tau_{\rm phys}} {\hbar} H_{\rm R}(x_i, p_i)\right)},}
\ee
and thus---up to some factor imposing energy conservation and a global phase $\exp(i\tau_{\rm phys} E)$ that has been dropped for clarity---one recovers the expected unitary time evolution in the emergent physical time (details are given below).

The identification of the suitable clock degree of freedom and its suitable boundary states has two associated effects.  On the one hand, the boundary states break the $N\to -N$ symmetry \eqref{sysy} of the configuration space amplitude \eqref{19} effectively unfreezing time evolution---whose dynamical form is that of the Schrödinger equation in physical time. On the other hand,  such symmetry breaking
resolves the apparent cosine tension by selecting the forward branch of the path integral (without the need of altering the time reflection symmetry of the fundamental dynamics). Explicitly, in the stationary phase approximation, the path integral representation of the previous amplitude becomes
\ba
\braket{\psi^{\rm out}_{\rm c};x^{\rm out}_i| x^{\rm in}_i; \psi^{\rm in}_{\rm c}}_{\rm phys.} \approx 
F_{\rm R}(\tau_{\rm phys}) \phantom{.}\exp\left( i \frac{S_{\rm R}[\tau_{\rm phys};x^{\rm out},x^{\rm in}]}{\hbar}\right).\ea
We will now prove these statements for some emblematic examples of clock systems: first for a harmonic oscillator clock (which is analytically simpler but has the drawback of being periodic), and then a free particle clock.
The conclusions should extend to general situations in the suitable corner of the system's Hilbert space where a clock variable is set in suitable good-clock states.

 \subsection{The harmonic oscillator clock}\label{HOC}

The simplest example will use a harmonic oscillator as a clock.  The use of coherent coherent states $\ket{z}$ for the clock states appear as the most natural choice (any squeezed state would also work). These have minimal uncertainties in their phase space variables which become negligible when $|z|\gg 1$. This is perhaps best illustrated in terms of their Wigner phase space distribution given by \cite{PhysRevA.48.2479}
\be W_{\ket{z}}(x,p)=\frac{2}{\pi} \exp\left(-2|\alpha(x,p)-z|^2\right)\ \ \ {\rm with} \ \ \ \alpha(x,p)\equiv \sqrt{\frac{m\omega}{2\hbar}} x  + \frac{i}{\sqrt{2\hbar m \omega}} p ,\ee
which defines a Gaussian distribution with the same features as the one for the ground state displaced from the complex phase space origin (in dimensionless coordinates) by the amount $z$.  
Coherent states form an over complete basis and have (normalized) overlaps $\braket{z|z'}/\sqrt{\braket{z|z}\braket{z'|z'}}\le \exp(-||z|-|z'||^2/2)$ that vanish exponentially if $z\not=z'$  and $(z,z')\to (\lambda z, \lambda z')$ for some $\lambda\gg 1$.
For these reasons, coherent states with $|z|\gg 1$ (large mean value of the occupation number) are the natural example of semiclassical states. 
These states are the candidates for good-clock states: their dynamics is well represented by the classical equations of motion of the harmonic oscillator with negligible uncertainties in the large occupation number limit $|z|\gg 1$.

For such a coherent state  $\ket{z}$ one has for instance
 \be
 \exp{(-iN H_c/\hbar)} \ket{z}=\ket{\exp{(-i\omega N)} z},
 \ee 
 where we are using the normal ordered Hamiltonian $H_c\equiv \hbar \omega (a^\dagger a)$ for simplicity, 
 with $a$ the annihilation operator, such that $a\ket{z}=z\ket{z}$.
 With all this, one can show that the clock amplitude in \eqref{ac} is given by
 \be
 A_c(N)=\frac{\braket{z_b|\exp{(-iN H_c/\hbar)}|z_a}}{\sqrt{\braket{z_b|z_b} \braket{z_a|z_a} }}= \frac{\braket{z_b|\exp{(-i\omega N)} z_a}}{\sqrt{\braket{z_b|z_b} \braket{z_a|z_a} }}, 
 \ee
so that using  $\braket{z'|z}=\exp(\bar z' z)$
we get
 \be
  A_c(N)=\exp\left(\frac{2\bar z_b z_a \exp(-i\omega N) -|z_a|^2-|z_b|^2}2\right).
 \ee
 Note that 
 \be
 |A_c(N)|
 =\exp\left(-\frac{|z_b-z_a \exp(-i\omega N)|^2}2\right)\le \exp\left(-\frac{||z_b|-|z_a||^2}2\right),
 \ee
 which implies that when $|z_a|\to \lambda |z_a|$ and $|z_b|\to \lambda |z_b|$, with $\lambda$ real and large,  the amplitude is exponentially suppressed in $\lambda$. As the semiclassical limit for the harmonic oscillator corresponds to large $|z|$ limit, the previous discussion imposes the requirement $|z_b|=|z_a|\equiv |z|$ for transition amplitudes not to vanish in the good-clock limit. Writing $z_a=|z| \exp(i\phi_a)$ and  $z_b=|z| \exp(i\phi_b)$ the clock amplitude becomes
 \be\label{aproxi}
 A_c(N)=\exp\left[{-|z|^2\left(1-\exp\left(i(\Delta\phi-\omega N)\right)\right)}\right],
 \ee
where $\Delta\phi\equiv \phi_a-\phi_b$. It is also important to note that $A_c(N=(\Delta\phi+2\pi n)/\omega )=1$ and  $\lim\limits_{|z|\to \infty} A_c(N\not=(\Delta\phi+2\pi n)/\omega)=0$.
 Thus, expanding around $N  =\tau_{\rm phys}$
  we get
 \be\label{clocko}
 A_c(N)=\exp\left[i |z|^2 \omega \left(\frac{\Delta\phi}{\omega}-N\right)-\frac{ |z|^2\omega^2}{2} \left(\frac{\Delta\phi}{\omega}-N\right)^2\right], 
 \ee
 which turns the integral over $N$  in \eqref{ac} into a simple Gaussian integral \footnote{\label{footo}The clock amplitude for a harmonic oscillator is periodic and the integral over $N$ would actually diverge. In order to regularize the integral  we neglected contributions from regions around $N=\tau_{\rm phys}+2\pi n/\omega$ for $n\not=0$. This amounts to assuming that the clock (as in practical cases) involves another degree of freedom that tracks the number of turns of the handles. With that assumption the Gaussian approximation around $N=\tau_{\rm phys}$ is consistent. It is also possible to consider the integral using the exact clock amplitude \eqref{aproxi} and a regularization that restricts the range of $x\equiv \omega(\tau_{\rm }-N)$ to the interval $[-\pi, \pi]$. The integral can be performed expanding the complicated exponential factor in \eqref{aproxi} in series, namely
 \be
 \exp\left[ |z|^2e^{ix} \right] = \sum_{n=0}^{\infty}  \frac{|z|^{2n}}{n!} e^{i nx}. 
 \ee Integrating each term of the series one finds that the amplitude can be written in terms of 
 exponentials and the lower incomplete gamma function. Using the known asymptotic behaviour
 of the latter \cite{PARIS2002323, Nemes:2015:RPIGII} in the good-clock limit $|z|^2 \gg 1$ the result \eqref{resorte} is recovered.}.
 Using that $E_c=|z|^2\hbar \omega$ is the mean value of the energy of the clock, and replacing the previous in \eqref{ac} we get
 \ba\label{resortes}
&& \braket{\psi^{\rm out}_{\rm c};x^{\rm out}_i| x^{\rm in}_i; \psi^{\rm in}_{\rm c}}_{\rm phys} =\bra{x^{\rm out}_i} \int \frac{dN}{t_p} \exp\left[i \frac{E_c}{\hbar} \left(\frac{\Delta\phi}{\omega}-N\right)-i \frac{N } {\hbar}(H_{\rm R}(x_i, p_i)-E) -\frac{E_c \omega}{2\hbar} \left(\frac{\Delta\phi}{\omega}-N\right)^2\right]\ket{x^{\rm in}_i} \n =\\
 &&= t_p^{-1} \exp\left({i \frac{E}{\hbar} \frac{\Delta\phi}{\omega}}\right) \bra{x^{\rm out}_i} { \sqrt{\frac{2\pi \hbar^2 |z|^2 }{E^2_c}}  \exp\left(-|z|^2 \frac{(H_{\rm R}(x_i, p_i)-E+E_c)^2}{2 E^2_c}\right)}  \exp\left({-i \frac{H_{\rm R}(x_i, p_i)}{\hbar} \frac{\Delta\phi}{\omega}}\right)\ket{x^{\rm in}_i}\n \\
 &&\approx  \frac{2\pi \hbar}{t_p}\,  \delta(H_{\rm R}(x_i, p_i)-E+E_c) \exp\left({i \frac{E}{\hbar} \frac{\Delta\phi}{\omega}}\right) \bra{x^{\rm out}_i} \exp\left({-i \frac{H_{\rm R}(x_i, p_i)}{\hbar} \frac{\Delta\phi}{\omega}}\right)\ket{x^{\rm in}_i},
\ea
where, in the last line, we have replaced the Gaussian factor by a Dirac delta distribution.  This is to be understood as follows.  In the good-clock limit $|z|\to \infty$, and assuming that $E_c$ remains much smaller than $E$, the Gaussian suppresses configurations where the energy differs from $E$ by an amount that is large in units of $E_c$. As the Gaussian with its prefactor defines a normalized measure one can represent it, under such circumstances, by a delta distribution.  Note that the last 
factor in the previous equation corresponds to the Schrödinger unitary evolution in physical time $\tau_{\rm phys}$, defined as
 \begin{equation}\label{titi}
  \tau_{\text{phys}}\equiv \frac{\Delta\phi}{\omega}.
\end{equation}
This matches the usual notion of time necessary to evolve the (complex) phase space point $z_a$ into $z_b$ for a classical harmonic oscillator: this is the clock time classically.
Accordingly,  we can write equation \eqref{resortes} as
\ba\label{resorte}
&& \braket{\psi^{\rm out}_{\rm c};x^{\rm out}_i| x^{\rm in}_i; \psi^{\rm in}_{\rm c}}_{\rm phys} \approx \n \\ && \ \ \ \ \ \ \ \ \ \ \ \ \approx  \frac{2\pi \hbar}{t_p}\,  \delta(H_{\rm R}(x_i, p_i)-E+E_c) \exp\left({i \frac{E}{\hbar}\tau_{\rm phys} }\right) \bra{x^{\rm out}_i} \exp\left({-i \frac{H_{\rm R}(x_i, p_i)}{\hbar} \tau_{\rm phys} }\right)\ket{x^{\rm in}_i}.
\ea
Thus, up to an unobservable global factor, we obtain the Schrödinger time evolution in physical time $\tau_{\rm phys}$. 

Using that $z=\sqrt{\frac{m\omega}{2\hbar}}\langle x \rangle + \frac{i}{\sqrt{2\hbar m \omega}} \langle p \rangle$ and $\phi = \arctan({\langle p \rangle}/({m\omega \langle x \rangle}))$, we get that 
$\Delta\phi^2=1/(4|z|^2)$. Thus
\begin{equation}
    \Delta \tau_{\text{phys}}^2 = \frac{\Delta\phi_a^2+\Delta\phi_b^2}{\omega^2}= \frac{1}{2\omega^2 \lvert z\lvert^2}.
    \label{eq:ClockError}
\end{equation}
Given that the characteristic time of the clock is $\tau_c\equiv 1/\omega$, we see that the good-clock regime $\Delta\tau_{\rm phys}/\tau_c=1/(2|z|)\ll 1$ is exactly the large occupation number $1\ll|z|$ limit anticipated as the {\em good-clock limit} at the beginning of the section.

\subsection{The free particle clock}

The free particle case is perhaps the simplest where the emergent time variable corresponds to monotonic parameter that avoids the limitation of the harmonic oscillator periodic time variable (artificially restricted, in the previous section, to the region $\omega N\in [-\pi, \pi]$---see footnote \ref{footo}). Thus, here we take $H_{\rm c}=p^2/(2 m)$ with $p$ and $m$ the momentum and mass of the free particle. Following the previous startegy, we focus our attention on a pair of `in' and `out' semiclassical boundary states for the particle degree of freedom given by 
\begin{eqnarray}\label{estados}
    \psi^{\rm in/out}(p)&=\left(\frac{\sigma^2}{\pi \hbar^2}\right)^{1/4} \exp\left(-\frac{\sigma^2}{2 \hbar^2} (p-{p_c})^2-\frac{i}{\hbar}{x_c^{\rm in/out}} (p-{p_c})\right).
\end{eqnarray}
The previous states are initially peaked at $x_c^{\rm in}$ with mean momentum $ p_c$ and finally peaked at $x_c^{\rm out}$ with mean momentum $p_c$ with $\Delta x=\sigma$ and $\Delta p=\hbar/\sigma$. Inspired by the result of the previous section, classical free propagation suggests that such a system, taken as a clock,  defines physical time via the relation
\be\label{weada}
\tau_{\rm phys}\equiv m\frac{x_b-x_a}{p_c}.
\ee 
Simple propagation of uncertainties associated with the intrinsic uncertainties of the basic variables in the states \eqref{estados},  $\Delta x=\sigma$ and $\Delta p=\hbar/\sigma$, leads to
\begin{equation}
    \frac{\Delta \tau_{\rm phys}}{\tau_{\rm phys}}=\sqrt{\frac{m\sigma^2}{2\tau_{\rm phys}^2{E}_c}+\frac{\hbar^2}{2{E}_cm\sigma^2}},
    \label{eq:quotient}
\end{equation}
where $E_c=p^2_c/(2m)$ is the mean energy of the clock.
The good-clock regime corresponds to $\Delta \tau_{\rm phys}/\tau_{\rm phys} \ll 1$, or equivalently
\begin{eqnarray}
    \frac{\hbar^2}{m\sigma^2 E_c} \equiv \epsilon_1\ll 1, \ \ \ \ \ \ 
       \frac{m\sigma^2}{{E}_c\tau^2_{\rm phys}} \equiv \epsilon_2\ll 1.
\end{eqnarray}
We will also need 
\be\frac{\hbar^2\tau^2_{\rm phys}}{m^2\sigma^4}= \frac{\epsilon_1}{\epsilon_2}\ll 1.\ee 
These conditions, defining the good-clock regime, are easy to interpret physically:
The first, $\epsilon_1\ll 1$, demands that the `zero-point-energy' $\Delta p^2/(2m)$ be smaller than the mean 
kinetic energy of the clock $E_c=p_c^2/(2m)$, or simply $\Delta p\ll p_c$. The second demands that the physical time $\tau_{\rm phys}$ be larger than the wave packet width traversing time $\Delta x/ v=m \Delta x/ p_c$.   The last condition can be written as
\be
\Delta v \tau_{\rm phys}\ll \Delta x.
\ee
The previous implies that the quantum spread of the wave function of the clock position does not grow more than the initial uncertainty in position.
Note that the good-clock limit is simpler in the case of the harmonic oscillator clock. This is due to the dynamical feature of harmonic-oscillator coherent states, for which the spread in phase-space variables around their classical trajectory are dynamically unchanged. The present clock model is, on the one hand, simpler because time is represented by a non periodic parameter; on the other hand, it is more involved when it comes to defining its good-clock limit. 

Now, we can consider the clock amplitude $A_c(N)$ defined in \eqref{ac} for boundary states as given in \eqref{estados}. Namely, 
\ba
A_c(N)&=& \left(\frac{\sigma^2}{\pi \hbar^2}\right)^{1/2} \int dp \,  \exp\left(-\frac{\sigma^2}{ \hbar^2} (p-{p_c})^2-\frac{i}{\hbar}{(x_c^{\rm in}-x_c^{\rm out })}(p-{p_c})- i \frac{N  }{\hbar} \frac {p^2}{2 m}\right)=\n \\
&=& \sqrt{\frac{\sigma^2}{\frac{iN  \hbar }{2m}+ {\sigma^2}}} \exp\left[-\frac{(N  -\tau_{\rm phys})^2}{\frac{\hbar^2 \tau^2_{\rm phys}}{2m\sigma^2 E_c} \left(\frac{N  }{\tau_{\rm phys}}\right)^2+\frac{2m \sigma^2}{E_c}}-i \frac{E_c N  }{\hbar}\left(\frac{1+\frac{\hbar^2\tau^2_{\rm phys}}{4m^2\sigma^4}\left(\frac{2N  }{\tau_{\rm phys}}-1\right)}{1+\frac{\hbar^2\tau^2_{\rm phys}}{4m^2\sigma^4} \left(\frac{N  }{\tau_{\rm phys}}\right)^2}\right)\right]\n \\
&=&\frac{1}{\sqrt{1+ \frac{i N}{2\tau_{\rm phys}}\sqrt{\frac{ \epsilon_1}{\epsilon_2}}}} \exp\left[-\frac{E_c}{2m\sigma^2} {\left({N  }-\tau_{\rm phys}\right)^2}-i \frac{E_c N  }{\hbar} +\sO\left(\frac{\epsilon_1}{\epsilon_2} \frac{N  }{\tau_{\rm phys}}\right)\right],\ea
%
%
%
%
where we have performed the Gaussian integral in $p$ exactly and used \eqref{weada} to express the result in terms of $\tau_{\rm phys}$. 
 Replacing the previous result in \eqref{ac} and performing the Gaussian integration in the lapse variable $N$ we get
 \ba\label{particula}
&& \braket{\psi^{\rm out}_{\rm c};x^{\rm out}_i| x^{\rm in}_i; \psi^{\rm in}_{\rm c}}_{\rm phys.} = \bra{x^{\rm out}_i} \int \frac{dN}{t_p} \exp\left[-\frac{E_c}{m\sigma^2} {\left({N  }-\tau_{\rm phys}\right)^2}-i \frac{E_c N  }{\hbar} -i \frac{N } {\hbar}(H_{\rm R}(x_i, p_i)-E) \right]\ket{x^{\rm in}_i} \n =\\
 &&=t_p^{-1}\exp\left({i \frac{E}{\hbar} \tau_{\rm phys}}\right) \bra{x^{\rm out}_i} { \sqrt{\frac{ 2 m\sigma^2\pi}{E_c}}  \exp\left(-\frac{(H_{\rm R}(x_i, p_i)-E+E_c)^2}{\frac{2 E_c\hbar^2}{m\sigma^2}}\right)}  \exp\left({-i \frac{H_{\rm R}(x_i, p_i)}{\hbar} \tau_{\rm phys}}\right)\ket{x^{\rm in}_i}\n \\
  &&\approx  \frac{2\pi \hbar}{t_p}\,  \delta(H_{\rm R}(x_i, p_i)-E+E_c) \exp\left({i \frac{E-E_c}{\hbar} \tau_{\rm phys}}\right) \bra{x^{\rm out}_i} \exp\left({-i \frac{H_{\rm R}(x_i, p_i)}{\hbar} \tau_{\rm phys}}\right)\ket{x^{\rm in}_i},
\ea
 where we have approximated the Gaussian in the second line by a Dirac delta taking the formal limit $\epsilon_1\to 0$ \footnote{Note that, considered as a distribution,  the formal limit of the Gaussian factor in equation \eqref{particula} when $\epsilon_1\to 0$ is 
 \be
\lim_{\epsilon_1 \to 0} { \sqrt{\frac{ 2\pi}{ \epsilon_1 E^2_c}}  \exp\left(-\frac{1}{2\epsilon_1}\frac{(H_{\rm R}(x_i, p_i)-E+E_c)^2}{E_c^2}\right)}=
2\pi \delta(H_{\rm R}(x_i, p_i)-E+E_c).
 \ee}.
Thus, up to an unobservable global factor, we obtain once more the Schrödinger time evolution in physical time $\tau_{\rm phys}$.

\subsection{Energy time uncertainty}

It is worth noting that a form of energy time uncertainty relation emerges naturally together with the notion of clock time. For the free particle clock we have---from the gaussian imposing energy conservation in \eqref{particula}---that $\Delta E ^2=E_c\hbar^2/(m\sigma^2)$. From 
\eqref{eq:quotient} it follows that $\Delta\tau_{\rm phys}^2=m\sigma^2/(2E_c) (1+\epsilon_1/\epsilon_2)$ so that
\be
\Delta\tau_{\rm phys}\Delta E =\frac{\hbar}{\sqrt{2}} \sqrt{1+\frac{\epsilon_1}{\epsilon_2}}>\frac{\hbar}{2}.
\ee 
For the harmonic oscillator one gets closer to the optimal. First equation \eqref{resorte} implies that $\Delta E^2=\hbar\omega E_c$.  Then, we recall that $\Delta \tau^2_{\rm phys}={1}/{(2 |z|^2\omega)}$ from \eqref{eq:ClockError}. The two together imply 
\be
\Delta\tau_{\rm phys}\Delta E =\frac{\hbar}{\sqrt{2}} >\frac{\hbar}{2},
\ee 
where we used that $E_c=|z| \hbar \omega$. In both the particle and harmonic oscillator clocks we do not get the optimal $\hbar/2$ 
result but instead ${\hbar}/{\sqrt{2}}$. This is due to an additional factor of 2 stemming from the double boundary condition in the path integral defining the clock states.

\subsection{Physical time evolution}
\vskip-.3cm
{\small \noindent{\em ``Traveling without moving.''}---David Lynch, {\em Dune} (1984).}
\vskip.3cm

Consider a given `initial' good-clock state, where initial is meant here as being the in-ket state $\ket{\psi_c^{\rm in}}$ in \eqref{pg}.
Then consider the physical state (solution of \eqref{concon})  $\ket{x_i^{\rm in}; \psi^0_{\rm c}}_{\rm phys}\equiv2\pi \delta(H-E)\ket{x_i^{\rm in}} \ket{\psi^{\rm 0}_{\rm c}} \in \sH_{\rm phys}$
which itself contains information in terms of Dirac observables or constants of motion. It is expected these constants of motion 
 to tell a story which in classical language would correspond to the history of the universe, given this particular physical state.
In the next Section we will exhibit concrete examples of such Dirac observables associated  to a harmonic oscillator subsystem and show that $x_s=\sqrt{\hbar/(2m_s\omega_s)} (a_s+a_s^\dagger)$ actually oscillates  in physical time $\tau$ when the expectation value is computed with different clock states (equation \eqref{amaca}).

Physical time $\tau$ evolution can also be extracted from the frozen all encompassing state of the universe $\ket{\Psi_{\rm U}}$ via the computation of (physical) transition amplitudes
(or shadows in the langange of \cite{Ashtekar:2002sn}) on the states $\bra{x^{\rm out}_i; \psi^{\rm out}_{\rm c}}$ for different clock states in the familly of good-clock states defined in previous sections. This produces  a collection of snap-shots of the history of the universe that can be organized in terms of the physical time $\tau$, schematically:      
\be
\begin{tikzpicture}[baseline=(A.base)]
  \node (A) at (-3,0) {$\ket{\Psi_{\rm U}}\equiv 2\pi \delta(H-E)\ket{\psi_{\rm R}}\ket{\psi^{\rm 0}_{\rm c}}\ \ \ \ \ \ \ \ \ \ \ \ \ \ \ \ \  \ \ \ \ \ \ \ \ \ \ \ \ \ \ \  {} $};
  \node (B1) at (3.5,1.5) {$\braket{\psi^{\rm III}_{\rm c} ; x^{\rm out}_i |x_i^{\rm in}; \psi^0_{\rm c}}_{\rm phys}=\braket{x^{\rm out}_i|\exp{\left(-i\tau^{\rm III} H_{\rm R}\right)} |x^{\rm in}_i}$}; 
  \node (B2) at (3.5,0) {$\braket{\psi^{\rm II}_{\rm c} ; x^{\rm out}_i |x_i^{\rm in}; \psi^0_{\rm c}}_{\rm phys}=\braket{x^{\rm out}_i|\exp{\left(-i\tau^{\rm II} H_{\rm R}\right)} |x^{\rm in}_i}$};
  \node (B3) at (3.5 ,-1.5) {$\braket{\psi^{\rm I}_{\rm c} ; x^{\rm out}_i |x_i^{\rm in}; \psi^0_{\rm c}}_{\rm phys} =\braket{x^{\rm out}_i|\exp{\left(-i\tau^{\rm I} H_{\rm R}\right)} |x^{\rm in}_i}$};

  \draw[->] (-2.5,0) -- (-0.5,1.5);
  \draw[->] (-2.5,0)  -- (B2);
  \draw[->] (-2.5,0) -- (-0.5,-1.5);
  
\node (C1) at (7.8,1.5) {${}$} ;
\node (C2) at (7.8,-1.5) {${}$};
\node (C3) at (8,0) {$\tau$};

  \draw[->, very  thick] (C2) -- (C1);
  
\end{tikzpicture}
\label{st},
\ee
which is the quantum version of spacetime evolution observables are replaced by quantum amplitudes---corresponding to the 
standard Schrödinger unitary evolution in our simple model---encoding the
quantum dynamical evolution in physical time. This captures the quantum analog of the classical spacetime picture
shown in Figure \ref{Tblem}.

This strengthens the view that the histories summed over in spin foams are not to be interpreted naively as quantum spacetime configurations \cite{Noui:2004iy}. Space time unfrozen information must be extracted from physical states when they are suitable---this is the case in our example, where the state of the universe $ \ket{\Psi_{\rm U}}=2\pi \delta(H-E)\ket{\psi_R} \ket{\psi^{\rm in}_{\rm c}}$ has a suitable initial arrow-of-time-selecting `initial' state. In general settings, a spacetime description will not be available and one would have to learn to extract physics without. This is not a problem as the terminology {\em the problem of time} or the {\em cosine problem} often suggests. 

In the context of classical gravity, we have learnt to accept the relatively limited usefulness of concepts like energy, energy conservation, etc, as  these notions are well defined only when very special conditions are given. Examples of this are  the analysis of test fields or test 
particles on given background geometries with symmetries (killing fields), or situations involving suitable boundary conditions (like in asymptotically flat spacetimes). It is perhaps possible that these circumstances could be recreated under suitable restrictions of the quantum states considered, and that a spacetime picture could be recovered in regions with sufficiently tamed quantum gravitational effects. Yet, quantum gravity  will remain timeless in its deepest quantum regime.

\section{The physical time is encoded in Dirac observables}\label{ddd}

In some conventional approaches mentioned in Section \ref{333}, time is not an allowed observable quantity according to Dirac's prescription. 
This is because the time variable thought of as a Hamiltonian vector field in the treatment reviewed in Section \ref{gaugefixing}, or as an operator in the one mentioned in Section \ref{Pauli}, does not preserve physical configurations: in both cases they are not Dirac observables (see \cite{Kuchar:1991qf} for a detailed discussion). The idea of evolving constants of motion \cite{PhysRevD.43.442} 
or the notion of partial observables \cite{Rovelli:2001bz} both lead to the construction of Dirac observables; however, issues of interpretation remain in that they continue to be labelled by kinematical quantities. This is at first sight what happens in our case where the projection into the physical Hilbert space of the kinematical states establishes a correlation between the clock reading $\tau_{\rm phys}$ and the other degrees of freedom (both of them {\em a priori} kinematical). Here we show explicitly that both the charaterization of the physical clock time  $\tau_{\rm phys}$, as well as the degrees of freedom of the rest of the universe, can indeed be encoded in Dirac observables. 

First, we note that most kinematical observables associated to the clock subsystem alone are generally trivial.  
For instance, the most general physical state of the closed system---a general state in $\sH_{\rm phys}$ annihilated by \eqref{concon}---can be expressed (in the energy basis) as
\be\label{previous}
\ket{\Psi}=\sum_{n,\alpha} \psi_{n,\alpha} \ket{n} \ket{E-n,\alpha}.
\ee
This implies that the reduced density matrix associated with the clock is
\be
\rho_c=\sum_n \rho_{nn} \ket{n} \bra{n},
\ee
i.e., diagonal in the energy basis, where $\rho_{nn} =\sum_\alpha |\psi_{n,\alpha}|^2$. This implies that 
$\braket{x_c} =\braket{p_c}=0$. This is due to the complete decoherence produced by the entanglement imposed by 
the constraint \eqref{concon}, or equivalently, by the fact that the `external time averaging' produced by the projection into the physical Hilbert space erases the pure gauge information contained in the canonical kinematical pair of the oscillator. Dirac observables 
(or gauge invariant observables) are generically `relational', in the sense that they involve correlations between a subsystem (here the clock) and the rest of the universe.  As such, the reduced density matrix where one traces out the rest of the universe can only be useful for computing 
`local' observables, which depend only on the degrees of freedom of the subsystem at hand. The only `local' Dirac  observable in our case---for which the expectation value computed with the reduced density matrix coincides with the one computed with the full one---is the energy of the clock, as a consequence of the lack of interactions assumed in equation \eqref{lack}. 

With the previous message in mind, let us write in more detail the physical state $\ket{\Psi_{\rm U}}$ in \eqref{st}, for the specially simple case of a harmonic oscillator clock. We will chose the clock state $\ket{\psi_{\rm c}}$ in a good-clock (semiclassical) state of the kind invoked in the previous section, namely,  $\ket{\psi_{\rm c}}=\ket{z}$.  Thus, we express the kinematical state in the energy eigen-basis as
\be
\ket{\psi_{\rm R}}\ket{\psi_{\rm c}}= \int d\sE\sum_{\alpha} \psi_R(\sE, \alpha) \ket{\sE,\alpha}\sum_{n=0}^\infty \frac{z^n}{n!} \ket{n},
\ee
where the clock has been put in a (kinematical) coherent state configuration with semiclassical parameters coded in $z$, and 
$\ket{\sE,\alpha}$ are eigenstates of the Hamiltonian of the rest of the universe, with  $\alpha$ is a quantum number labelling the degeneracy of the energy levels $\sE$ (we assume for notational simplicity that $\alpha$ is discrete).
The physical state of the universe is constructed by projecting the previous on the solutions of \eqref{concon}, which explicitly becomes
\ba
\ket{\Psi_{\rm U}}&=&\frac{2\pi \hbar}{t_p}\delta(H-E)\ket{\psi_{\rm R}}\ket{\psi_{\rm c}} \n \\
&=&\frac{2\pi \hbar}{t_p} \delta(H-E) \int d\sE\sum_{\alpha} \psi_R(\sE, \alpha) \ket{\sE,\alpha}\sum_{n=0}^\infty \frac{z^n}{n!} \ket{n}
\n \\
&=& \frac{2\pi \hbar}{t_p} \int d\sE \sum_{n=0}^\infty \sum_{\alpha} \psi_R(\sE, \alpha) \ket{\sE,\alpha} \frac{z^n}{\sqrt{n!}} \ket{n}\delta(\hbar \omega  n +\sE -E)
\n \\
&=&\frac{2\pi \hbar}{t_p} \sum_{n=0}^{{\rm Int}\left[\frac{E}{\hbar\omega}\right]}  \sum_{\alpha} \psi_R(E-\hbar \omega n, \alpha) \ket{E-\hbar \omega n,\alpha} \frac{z^n}{\sqrt{n!}} \ket{n}.
\ea
The symbol ${\rm Int}\left[\ \ \right]$ denotes the integer part of its argument. The previous physical state---which is clearly in the kernel of the Hamiltonian constraint \eqref{concon}---exhibits the expected entanglement between the clock degree of freedom (here a harmonic oscillator) and the rest of the universe.

The physical state inherits from the kinematical oscillator coherent state of the clock some features. 
As explained in the introduction, we will not be concerned with the hard questions associated with the interpretation of quantum theory of a closed universe. Nevertheless, we can  still ask the question of whether there are Dirac observables---operators commuting with the Hamiltonian constraint \eqref{concon}---that contain the information about the state of the clock. 
In order to address this, it is convenient to introduce the notation 
\be
|n,\alpha)\equiv \ket{n}\ket{E-\hbar \omega n,\alpha}. 
\ee
The previous defines an element of an orthonormal basis of the physical Hilbert space (the kernel of $C$ introduced in \eqref{concon}). Now we introduce
physical creation and annihilation operators, $a_{\rm phys}$ and $a_{\rm phys}^\dagger$,  defining their action on the previous basis as follows \footnote{In the explicit example of \cite{Rovelli:1990jm},  where the rest of the universe is another harmonic oscillator, 
one can explicitly define $$ {a_{\rm phys}}\equiv \left(\frac1 {\sqrt{E-a^\dagger_c a_c}}\right){a}_c  \otimes {a}^{\dagger}_R\ \ \ {\rm with}\ \ \ {a_{\rm phys}^\dagger}\equiv {a}_c^\dagger \left(\frac1 {\sqrt{E-a^\dagger_c a_c}}\right) \otimes {a}^{\dagger}_R, 
$$ where $a_c$ is the annihilation operator of the clock oscillator and $a_{R}$ the one corresponding to the additional harmonic oscillator idealized rest-of-the universe degree of freedom. In the general case, as long as the spectrum of the Hamiltonian of the rest of the universe is known, an explicit expression for these Dirac observables can be given based on the action \eqref{seventyeight}. Unlike in  \cite{Rovelli:1990jm} or the case of evolving constants of motion \cite{Ashtekar1991-ASHCPO, PhysRevD.43.442, Rovelli:1990jm}, the observables in this Section are 
defined directly at the quantum level. }:
\be\label{seventyeight}
a_{\rm phys}|n,\alpha)=\sqrt{n}|n-1,\alpha)\ \ \ {\rm and} \ \ \ a^\dagger_{\rm phys}|n,\alpha)=\sqrt{n+1}|n+1,\alpha).
\ee 
The previous are Dirac operators---$ [a_{\rm phys} , C]=[a^\dagger_{\rm phys}, C]=0$---,hermitian conjugate of each other, satisfying the standard  commutation relations
$[a_{\rm phys} ,a^\dagger_{\rm phys}]=1$. These operators are `non-local' in the sense of the previous discussion as they preserve the entanglement between the clock and the rest of the universe (imposed by \eqref{concon}). For these operators we have
\be 
\frac{\braket{\Psi_{\rm U}|a_{\rm phys}| \Psi_{\rm U}}}{\braket{\Psi_{\rm U} |\Psi_{\rm U}}}=z \left(\frac{ \sum\limits_{n=0}^{{\rm Int}\left[\frac{E}{\hbar\omega}-1\right]}  \frac{|z|^{2n}}{{n!}}} { \sum\limits_{n=0}^{{\rm Int}\left[\frac{E}{\hbar\omega}\right]}\frac{|z|^{2n}}{{n!}} }\right)\approx z,
\ee
where the last approximate sign denotes the fact that this is an equality in the $E/(\hbar \omega)\gg1$ regime. Therefore, the semiclassical parameters of the clock system---the information of the clock handles, encoded in the phase of $z$ (recall \eqref{titi})---are captured by the expectation value of the previous Dirac observables, which give an operational physical meaning to the variable $\tau_{phys}$ in the previous discussion (which is just a function of $z$). At the physical level such `non-locality'  can be interpreted by saying that `observing' $\tau_{\rm phys}$ from within the closed universe necessarily involves an energy exchange between the clock and the `observer' (who is part of the rest of the universe), in such a way that the total energy remains $E$. Thus, our prescription characterizes physical clock time relationally and explicitly in a way that is associated with genuine Dirac observables.

One can also express the motion of a subsystem in terms of the clock variable in an explicit manner.
For that, let us assume that the subsystem is another harmonic oscillator with frequency $\omega_s$, and write the elements of a basis of the physical Hilbert space as
\be
\ket{n,m,E-n\hbar\omega_c-m\hbar\omega_s, \alpha}=\ket{n} \ket{m} \ket{E-\hbar \omega n,\alpha},
\ee
where the first entry is the clock occupation number, the second that of the new harmonic oscillator (the subsystem), and the last two  are the energy and degeneracy of the  rest of the universe.
The annihilation operator of the clock is the previously defined Dirac operators $a_c$  
\be
a_c\ket{n,m,E-n\hbar\omega_c-m\hbar\omega_s, \alpha}=\sqrt{n} \ket{n-1,m,E-(n+1)\hbar\omega_c-m\hbar\omega_s, \alpha},
\ee 
while a second Dirac operator $a_s$, characterizing the oscillator subsystem, is defined as
\be
a_s \ket{n,m,E-n\hbar\omega_c-m\hbar\omega_s, \alpha}=\sqrt{m} \ket{n+k_s,m-1,E-n\hbar\omega_c-m\hbar\omega_s+\varepsilon_s, \alpha},\ee 
where 
\be
k_s\equiv {\rm Int}\left[\frac{\omega_s}{\omega_c}\right]\ \ \ {\rm and} \ \ \ \varepsilon_k=\hbar\omega_s\left(\frac{\omega_s}{\omega_c}-k_s\right),
\ee
As before, both $a_c$ and $a_s$ are Dirac operators---as $[C,a_c]=[C,a_s]=0$ with $H_c=\hbar \omega_c a_c^\dagger a_c$ and $H_s=\hbar \omega_s a_s^\dagger a_s$ and $[H_c,H_s]=0$. However, we have that 
\be
[a_c, a_s] =\left(\sqrt{N_c+1}-\sqrt{N_c}\right) a_s^{(0)},
\ee
with $a_s^{(0)}\ket{n,m,E-n\hbar\omega_c-m\hbar\omega_s, \alpha}=\sqrt{m} \ket{n,m-1,E-n\hbar\omega_c-(m+1)\hbar\omega_s, \alpha}$.
Thus, $[a_c, a_s]=\sO(1/\sqrt{N_c})$, implying, that for the states involved in the good-clock limit,  the two operators commute (in the sense that the non commutativity is lost in the classical limit of the clock state). One can show that
\be
[a_c^\dagger a_c,a_s]= k_s a_s,
\ee
which implies that $a_s(\tau_c)\approx \exp(-i\omega_s\tau_c) a_s$, or (alternatively)
\be\label{amaca}
 \frac{\braket{\Psi_1|a_s| \Psi_1}}{\braket{\Psi_2|a_s| \Psi_2}}\approx \exp(-i\omega_s\tau_c^{1\to 2}), 
\ee
where the approximation is valid in the good-clock limit $|z|\gg1$, and $E \gg |z_c|^2+|z_s|^2$.

\section{Why the Cosine Problem Is Not a Problem}

Coming back to the expression of the transition amplitudes in the path integral representation---as stated in equation \eqref{coscos}---we have
 \be\label{pipo}
{\braket{x^{\rm out}_i| x^{\rm in}_i}_{\rm phys.}={\hbar}\int \frac{dN}{t_p} \left(|F[N]|  \cos\left({\frac{1}{\hbar}S\left[N , x^{\va \rm cl}_i(t)\right]}+n\frac{\pi}{4}\right)\right)}, 
\ee
which is the universal form of the path integral for timeless systems. The previous matches the formal structure of the path integral of 
generally covariant systems such as general relativity as explicitly shown in \eqref{cosy}. As mentioned in Section \ref{coseno},  this is precisely the general structure found in certain path integral
definitions of quantum gravity in three dimensions as well as for models of quantum gravity in four dimensions. 
If among the $x_i$ one identifies a clock system---with total Hamiltonian $H=H_c+H_R$, where  $H_c$ and $H_R$ represent the clock and the rest-of-the-universe Hamiltonian---then we have
\be
S\left[N , x^{\va \rm cl}_i(t)\right]=S_c\left[N , x_c\right]+S_R\left[N , x^{\va \rm cl}_{i\not=c}(t)\right],
\ee
and thus \eqref{pipo} becomes 
\ba
&& \braket{\psi^{\rm out}_{\rm c};x^{\rm out}_i| x^{\rm in}_i; \psi^{\rm in}_{\rm c}}_{\rm phys}=  \n \\ 
&& \n \int {dN} \left(\overbrace{\frac{1}{t_p}\left[\int  \overline{\psi}(x_c^{\rm out})\psi(x_c^{\rm in}) |F_c[N]| \exp\left({\frac{i}{\hbar}S\left[N , x^{\va \rm cl}_{c}(t)\right]}+\frac{\pi}{4}\right) dx_c^{\rm out} dx_c^{\rm in} \right]}^{A_c(N)} |F_R[N]|  \exp\left({\frac{i}{\hbar}S\left[N , x^{\va \rm cl}_{i\not=c}(t)\right]}+(n-1)\frac{\pi}{4}\right) \right)+\\
&& \n +\int {dN} \left(A_c(-N) |F_R[-N]|  \exp\left({\frac{i}{\hbar}S\left[-N , x^{\va \rm cl}_{i\not=c}(t)\right]}+(n-1)\frac{\pi}{4}\right) \right),
\ea
where in the last line we have replaced directly the clock amplitude $A_c(-N)$ in order to shorten the expression. The explicit evaluation of  $A_c(N)$  was performed in the good-clock limit for two emblematic examples:  the harmonic oscillator and the free particle clock. One can synthetically represent the result of Section \ref{calculos}---equations \eqref{resorte} and \eqref{particula}---by the following identification (where a dynamically irrelevant global phase has been ignored) 
\be
A_c(N)\ \ \ \ \ \ \longrightarrow_{{}_{{}_{{}_{\!\!\!\!\!\!\!\!\!\!\!\!\!\!\!\!\!\!\!\!\!\!\!\!\!\!\!\!  \rm good\ clock\ limit}}}}  \frac{2\pi \hbar}{t_p}\,  \delta(H_{\rm R}(x_i, p_i)-E+E_c)  \delta(N-\tau_{\rm phys}),
\ee
explicitly showing that the clock (semiclassical) state is the one responsible of selecting the forward propagation branch of the cosine yielding 
\be\label{pg}
\braket{\psi^{\rm out}_{\rm c};x^{\rm out}_i| x^{\rm in}_i; \psi^{\rm in}_{\rm c}}_{\rm phys}
=\frac{2\pi \hbar}{t_p}\left(|F_R[\tau_{\rm phys}]|  \exp\left({\frac{i}{\hbar}S_R\left[\tau_{\rm phys} , x^{\va \rm cl}_{i\not=c}(t)\right]}+(n-1)\frac{\pi}{4}\right)\right)  \delta(H_{\rm R}(x_i, p_i)-E+E_c) .
\ee
The factor $ \delta(H_{\rm R}-E+E_c) $ imposes the energy conservation for the degrees of freedom in the rest of the universe. 
Note that, as shown explicitly in the discussion at the end of Section \ref{ddd}, if one would focus on a small part of the rest of the universe that we can call the system of interest---assuming in turn that $H_R=H_{\rm system}+H_{R'}$ and than the energy $E-E_c$ is much larger that any possible practical value of the energy of the system---then 
one can forget (as far as the system is concerned) the restriction imposed by  the factor $ \delta(H_{\rm system}+H_{\rm R'}-E+E_c)$
on the states of the system.  In this way one recovers, from a 
fundamentally timeless quantum theory, the standard amplitudes (customarily encountered in quantum mechanics)
with time evolution in $\tau_{\rm phys}$ emerging when an internal clock degree of freedom is in a suitable semiclassical configuration.

Even when the toy model upon which we base our analysis is quite simple, its implications for the spin foam program 
of non-perturbative gravity is insightful. On the one hand, it shows that the  {\em cosine problem} is not an issue 
that has to be dealt by modifying in any way the definition of the transition amplitudes: projection into the kernel of the quantum constraints 
requires integration over both positive and negative values of the lapse function. Proposals inspired by the formal treatment of the 
rather influential reference \cite{Teitelboim:1981ua}, aiming at producing a seemingly consistent amplitude with forward propagation run into conflict with the imposition of the quantum constraints and thus general covariance. On the other hand, the previous discussion---and the conclusion drawn from \eqref{st}---also illustrates how a spacetime interpretation of the path integral of gravity can only emerge under very special circumstances when an appropriate good-clock variable can be identified (which of course in the case of gravity  will have to correspond to some field theoretic degree of freedom in a suitable semiclassical state).

\section{Discussion}\label{discu}

We have shown that the presence of a clock degree of freedom in a suitable semiclassical state, that we 
refer to  as the {\em good-clock state},  has the effect of selecting the forward propagating amplitude in the path integral representation of an otherwise timeless system. This was explicitly illustrated here for a toy model defined by a non relativistic quantum mechanical closed system with fixed total energy eigenvalue $E$. These results suggest clearly what we should expect for path integral amplitudes of background independent theories in view of  the {\em cosine problem}.

Despite the simplicity of the model, it captures the conceptual features of generally covariant systems. Thus, even when it does not solve all issues related to understanding dynamics in quantum gravity, this simple example provides an insightful perspective where time evolution is only emergent if suitable degrees of freedom happen to be in suitable {\em good-clock} states in an otherwise `frozen' timeless fundamental structure.

Such suitable {\em good-clock} states select an arrow of time. In practice, even though they are very special, such states are expected to capture realistic features of nature due to the particular form of the state of the universe around us, which---for reasons that remain fundamentally unclear---is time asymmetric, as most clearly illustrated by the validity of the second law of thermodynamics \cite{penrose1979singularities}. This universal arrow of time manifests itself in the existence around us of a vast set of {\em good-clock} states, which have made us accustomed to describing physics in terms of a classical parameter we call time, with evolution governed by forward unitary Schrödinger dynamics.

From this perspective, what is usually referred to as the problem of time is simply a reflection of the fundamental nature of genuinely isolated or closed systems—of which generally covariant theories such as general relativity are the quintessential example. This is just the way fundamental dynamics is, and not a problem to be solved. Time evolution with respect to a clock can only be recovered under very special conditions. There are many reasons to expect that such conditions will not apply under extreme regimes of strong quantum gravitational effects (such as near the initial big bang singularity in cosmology or close to singularities inside black holes). In those cases, physics remains timeless, and we will have to learn how to extract physical information without relying on any notion of time evolution.

We would like to emphasize the explicit picture of the emergence of time in our simplified (non-relativistic) closed system of Section \ref{ddd}, where we are able to characterize dynamical information entirely in terms of Dirac (energy-conserving or gauge-invariant) observables. This provides an explicit realization of the idea of evolving constants of motion \cite{PhysRevD.43.442}, in a setting where the data that label them and give them physical meaning are genuinely gauge-invariant notions defined relationally from within the idealized closed system, and formulated directly in the quantum framework.

\section{Acknowledgments} 

We are grateful to Pietro Doná and Daniel Sudarsky for helpful exchanges. This work was made possible through the support of the WOST, WithOut SpaceTime project (https://withoutspacetime.org), supported by Grant ID63683 from the John Templeton Foundation (JTF). The opinions expressed in this work are those of the author(s) and do not necessarily reflect the views of the John Templeton Foundation.

\begin{appendix}
\clearpage

\end{appendix}

\providecommand{\href}[2]{#2}\begingroup\raggedright\endgroup

%

\end{document}